\documentclass{aa}

\newcommand{\p}{\tiny}
\newcommand{\s}{\scriptsize}
\newcommand{\acena}{$\alpha$\,Cen\,A}
\newcommand{\acenb}{$\alpha$\,Cen\,B}

\newcommand{\Wlam}   {$W_{\lambda}$}
\newcommand{\Wlams}  {$W_{\lambda}$s}
\newcommand{\Wlamb}  {$W_{\lambda}$~}
\newcommand{\Wlamsb} {$W_{\lambda}$s~}
\newcommand{\Teff}   {$T_{\rm eff}$}
\newcommand{\Teffs}  {$T_{\rm eff}$s}
\newcommand{\Teffb}  {$T_{\rm eff}$~}
\newcommand{\Teffsb} {$T_{\rm eff}$s~}
\newcommand{\tpm}[2]{#1{\tiny$\pm$#2}}
\newcommand{\mezero}{\textcolor{white}{$+$}0.00}

\usepackage{graphicx}
\usepackage{multirow}
\usepackage{color}

\begin{document}

\title{The Alpha Centauri binary system
\thanks{Based on observations collected at Observat\'orio do Pico dos Dias (OPD),
operated by the Laborat\'orio Nacional de Astrof\'{\i}sica, CNPq, Brazil.}
\subtitle{Atmospheric parameters and element abundances}
}

\author{G. F. Porto de Mello, W. Lyra\fnmsep\thanks{Present address: Department of Physics and Astronomy,
Uppsala Astronomical Observatory, Box 515, 751\,20 Uppsala,
Sweden},~
  and G. R. Keller\fnmsep\thanks{Present address: Universidade de S\~ao
Paulo, Instituto de Astronomia, Geof\'isica e Ci\^encias
Atmosf\'ericas, Depto. de Astronomia, Rua do Mat\~ao 1226, S\~ao
Paulo-SP 05508-900, Brazil}}

\offprints{gustavo@ov.ufrj.br}

\institute{Observat\'orio do Valongo, Universidade Federal do Rio de Janeiro, Ladeira do Pedro Ant\^onio, 43\\
20080-090 Rio de Janeiro, RJ, Brazil
}

\date{Received, accepted}

\authorrunning{Porto de Mello, Lyra \& Keller}
\titlerunning{Abundance analysis of $\alpha$\,Cen\,A and B}

\abstract
{The $\alpha$ Centauri binary system, owing to its duplicity,
proximity and brightness, and its components' likeness to the Sun,
is a fundamental calibrating object for the theory of stellar
structure and evolution and the determination of stellar
atmospheric parameters. This role, however, is hindered by a
considerable disagreement in the published analyses of its
atmospheric parameters and abundances.}
{We report a new spectroscopic analysis of both components of
the $\alpha$ Centauri binary system, compare published analyses
of the system, and attempt to quantify the discrepancies still
extant in the determinations of the atmospheric parameters and
abundances of these stars.}
{The analysis is differential with respect to the Sun, based
on spectra with R = 35\,000 and signal-to-noise ratio $\geq$
1\,000, and employed spectroscopic and photometric methods to
obtain as many independent \Teffb determinations as possible. The
atmospheric parameters are also checked for consistency against
the results of the dynamical analysis and the positions of the
components in a theoretical HR diagram.}
{The spectroscopic atmospheric parameters of the system are found
to be {\Teffb = (5847 $\pm$ 27) K, $[$Fe/H$]$ = $+$0.24 $\pm$ 0.03,
$\log g$ = 4.34 $\pm$ 0.12 and $\xi_t$ = 1.46 $\pm$ 0.03
km\,s$^{-1}$, for $\alpha$\,Cen\,A, and \Teffb = (5316 $\pm$ 28) K,
$[$Fe/H$]$ = $+$0.25 $\pm$ 0.04, $\log g$ = 4.44 $\pm$ 0.15 and
$\xi_t$ = 1.28 $\pm$ 0.15 km\,s$^{\rm -1}$ for $\alpha$\,Cen\,B}.
The parameters were derived from the simultaneous excitation \&
ionization equilibria of \ion{Fe}{i} and \ion{Fe}{ii} lines.
\Teffsb were also obtained by fitting theoretical profiles to the
H$\alpha$ line and from photometric calibrations.}
{Good agreement was reached between the three criteria for
$\alpha$\,Cen\,A. For $\alpha$\,Cen\,B the spectroscopic \Teffb is
$\sim$140 K higher than the other two determinations. We discuss
possible origins of this inconsistency, concluding that the
presence of NLTE effects is a probable candidate, but we note that
there is as yet no consensus on the existence and cause of an
offset between the spectroscopic and photometric \Teffb scales of
cool dwarfs. The spectroscopic surface gravities also agree with
those derived from directly measured masses and radii. An average
of three independent \Teffb criteria leads to \Teff (A) = (5824
$\pm$ 26) K and \Teff (B) = (5223 $\pm$ 62) K. The abundances of
Na, Mg, Si, Mn, Co and Ni and, possibly, Cu are significantly
enriched in the system, which also seems to be deficient in Y and
Ba. This abundance pattern can be deemed normal in the context of
recent data on metal-rich stars. The position of $\alpha$\,Cen\,A
in an up-to-date theoretical evolutionary diagrams yields a good
match of the evolutionary mass and age (in the 4.5 to 5.3 Gyr
range) of with those from the dynamical solution and seismology,
but only marginal agreement for $\alpha$\,Cen\,B, taking into
account its more uncertain \Teff.}


\maketitle

\section{Introduction}

The $\alpha$ Centauri binary system, composed of two solar-type
stars (HD 128620 and 128621), is one of the brightest in the sky
and figures as our second closest galactic neighbor, 1.34 parsec
away. The star closest to the Sun is the M5.5 dwarf Proxima
Centauri (Gliese \& Jahreiss 1991), $\sim$15,000 A.U. away from
the $\alpha$ Centauri binary, and its gravitational connection to
the system is still a topic of controversy. Anosova et al. (1994)
proposed that Proxima has a hyperbolic orbit around the inner
pair, and that the three stars might form part of a more extended
kinematical group. Wertheimer \& Laughlin (2006), however, found
the distance between Proxima and the pair as comparable to the
Hill radius of the latter, whereby the Galactic potential becomes
dominant over that of the inner pair and the system becomes
unbound. These authors favor the existence of a physically bound
triple system, suggesting that Proxima is presently at the
apoastron of its orbit. Highly precise monitoring of radial
velocity variations of the system by Endl et al. (2001) constrains
the upper limit of the mass of putative planetary or substellar
companions of the system as less than 3.5 Jupiter masses (actually
less than one Saturn mass if coplanar orbits are assumed).

The proximity of the $\alpha$ Centauri system provides a well
determined parallax, and its brightness allows the acquisition of
extremely high-quality spectra. Moreover, its binary nature and
relatively short period of 80 years enables the hypothesis-free
accurate determination of masses (Pourbaix et al. 1999, 2002). If
we couple to these facts their being very solar-like, the $\alpha$
Centauri stars thus appear as objects of fundamental importance in
the calibration of evolutionary tracks, theoretical isochrones and
model atmospheres, hence the great interest in the precise
determination of their atmospheric parameters, evolutionary state
and chemical composition.

\begin{table*}
\caption[]{A non-exhaustive review of determinations of effective
temperatures \Teff, iron abundances [Fe/H], logarithm of surface
gravity $\log g$ and microturbulence velocities $\xi_t$ for
$\alpha$\,Cen\, A and B. The notation used from columns 5 to 7
stands for: $a$ - Excitation equilibrium; $b$ - Wings of Balmer
lines; $c$ - Trigonometric parallax; $d$ - Wings of strong lines;
$e$ - Ionization equilibrium; $f$ - Photometric color indexes;
$direct$ - directly measured luminosity, mass and radius. Note
that, generally, the microturbulence velocities have not the same
zero point and cannot be directly compared. Typical errors,
respectively, in \Teff, log g and [Fe/H] are $\leq$ 100 K, $\leq$
0.2 dex and $\leq$ 0.1 dex, but note that these estimates usually
do not include systematic uncertainties, and that not all authors
provide error determinations, neither labor under the same
definitions for them.} \label{hist-review}
\begin{center}
\begin{tabular}{lcc ccc cc}\hline
\multicolumn{8}{c}{$\alpha$\,Centauri\,A}\\\cline{1-8}
Reference&\multicolumn{4}{c}{Atmospheric
Parameter}&&\multicolumn{2}{c}{Method
Used}\\\cline{2-5}\cline{7-8}

&\Teff(K)&$\log g$&$\xi_t$(km\,s$^{-1}$)&$\rm [Fe/H]$
&&\Teff&$\log g$\\\hline
French \& Powell (1971)            & 5770 &   -  &  -   &$+$0.22&&$a$  &-   \\
Soderblom (1986)                   & 5770 &   -  &  -   &  -    &&$b$  &-   \\
England (1980)                     & 5750 & 4.38 & 1.0  &$+$0.28&&$b,c,d$&$c,d$\\
Bessell (1981)                     & 5820 & 4.25 & 1.7  &$-$0.01&&$a,e$ &$e$ \\
Smith et al. (1986)                & 5820 & 4.40 & 1.54 &$+$0.20&&$a,e$ &$d$ \\
Gratton \& Sneden (1987)           & 5750 & 4.38 & 1.2  &$+$0.11&&$f$  &$c,e$\\
Abia et al. (1988)                 & 5770 & 4.5  & 1.0  &$+$0.22&&$b$  &$f$ \\
Edvardsson (1988)                  & -    & 4.42 & -    &$+$0.28&&-    &$d$ \\
Furenlid \& Meylan (1990)          & 5710 & 4.27 & 1.0  &$+$0.12&&$a,e$ &$e$ \\
Chmielewski et al. (1992)          & 5800 & 4.31 & -    &$+$0.22&&$b$  &$c$ \\
Neuforge-Verheecke \& Magain (1997)& 5830 & 4.34 & 1.09 &$+$0.25&&$a$  &$ce$\\
Allende-Prieto et al. (2004)       & 5519 & 4.26 & 1.04 &$+$0.12&&$f$  &$c$ \\
Doyle et al. (2005)                & 5784 & 4.28 & 1.08 &$+$0.12&&$direct$  &$direct$ \\
Santos et al. (2005)               & 5844 & 4.30 & 1.18 &$+$0.28&&$a$  &$a$ \\
del Peloso et al. (2005a)          & 5813 & 4.30 & 1.23 &$+$0.26&&$b,f$&$c$ \\
Valenti \& Fischer (2005)          & 5802 & 4.33 & -    &$+$0.23&&$a$  &$e$ \\
This work                          & 5824 & 4.34 & 1.46 &$+$0.24&&$a,b,e,f$&$c,e$ \\
\\\hline
                                   &      &      &      &       &&     &    \\\hline

\multicolumn{8}{c}{$\alpha$\,Centauri\,B}\\\cline{1-8}
Reference&\multicolumn{4}{c}{Atmospheric
Parameter}&&\multicolumn{2}{c}{Method
Used}\\\cline{2-5}\cline{7-8} &\Teff(K)&$\log
g$&$\xi_t$(km\,s$^{-1}$)&$\rm [Fe/H]$&&\Teff&$\log
g$\\\hline
French \& Powell (1971)             & 5340 &  -   & -    &$+$0.12&&$a$  &-   \\
Soderblom (1986)                    & 5350 &  -   & -    &  -    &&$b$  &-   \\
England (1980)                      & 5260 & 4.73 & 1.1  &$+$0.38&&$b,c,e$&$c,e$\\
Bessell (1981)                      & 5350 & 4.5  & 1.0  &$-$0.05&&$a,e$ &$e$ \\
Smith et al. (1986)                 & 5280 & 4.65 & 1.35 &$+$0.20&&$a,e$ &$d$ \\
Gratton \& Sneden (1987)            & 5250 & 4.50 & 1.0  &$+$0.08&&$f$  &$c,e$\\
Abia et al. (1988)                  & 5300 & 4.5  & 1.5  &$+$0.14&&$b$  &$f$ \\
Edvardsson (1988)                   & -    & 4.65 & -    &$+$0.32&&-    &$d$ \\
Chmielewski et al. (1992)           & 5325 & 4.58 & -    &$+$0.26&&$b$  &$c$ \\
Neuforge-Verheecke \& Magain (1997) & 5255 & 4.51 & 1.00 &$+$0.24&&$a$  &$c,e$\\
Allende-Prieto et al. (2004)        & 4970 & 4.59 & 0.81 &$+$0.18&&$f$  &$c$ \\
Santos et al. (2005)                & 5199 & 4.37 & 1.05 &$+$0.19&&$a$  &$a$ \\
Valenti \& Fischer (2005)           & 5178 & 4.56 & -    &$+$0.22&&$a$  &$e$ \\
This work                           & 5223 & 4.44 & 1.28 &$+$0.25&&$a,b,e,f$ &$c,e$ \\
\hline
\end{tabular}
\end{center}
\end{table*}

The brightness of the system's components also favor the
determination of internal structure and state of evolution by
seismological observations. The analysis of the frequency spectrum
and amplitudes of both photometric and spectroscopic oscillations
in the outer layers of solar-type stars, driven by convection, can
yield otherwise unobtainable information on internal structure,
such as the depth of the convection zone and the density and
temperature profiles. They can also provide independent checks on
stellar masses, ages and chemical composition. Y{\i}ld{\i}z (2007),
Eggenberger et al. (2004) and Thoul et al. (2003) have agreed on
an age for the system between 5.6 and 6.5 Gyr. Miglio \&
Montalb\'an (2005) propose model-dependent ages in the 5.2 to 7.1
Gyr interval. However, they also note that fixing the non-seismic
observables, namely masses and radii, leads to an age as large as
8.9 Gyr, proposing that further seismological observations may be
needed to clarify this apparent discrepancy between the
independent observation of the oscillation spectra and the
directly measured masses and radii. The previous analysis of
Guenther \& Demarque (2000) favors a slightly higher age of
$\sim$7.6 Gyr. The masses are very well constrained at M$_{\rm A}$
= 1.105 $\pm$ 0.007 and M$_{\rm B}$ = 0.934 $\pm$ 0.006 in solar
masses (Pourbaix et al. 2002), which, along with
interferometrically measured (in solar units) radii of R$_{\rm A}$
= 1.224 $\pm$ 0.003 and R$_{\rm B}$ = 0.863 $\pm$ 0.005 (Kervella
et al. 2003) yield surface gravities (in c.g.s. units) of log
g$_{\rm A}$ = 4.307 $\pm$ 0.005 and log g$_{\rm B}$ = 4.538 $\pm$
0.008, an accuracy seldom enjoyed by stellar spectroscopists.
Altogether, these data pose very tight constrains on the modelling
of fundamental quantities of internal structure, such as
mixing-length parameters and convection zone depths.

Nevertheless, the state of our current understanding of this
system still lags behind its importance, since published
spectroscopic analyses reveal considerable disagreement
(Table~\ref{hist-review}) in the determination of atmospheric
parameters and chemical abundances, particularly for component B,
though most authors agree that the system is significantly
metal-rich with respect to the Sun. This fact is embarrassing,
even in our modern era of massive surveys, since the individual
study of key objects is necessary to quantify systematic errors
which might be lurking inside huge databases, and cannot be
reduced with large number statistics. Indeed, considering only
those analyses since the 90s, eight performed a detailed analysis
of the atmospheric parameters and chemical composition of
$\alpha$\,Cen\,A (Furenlid \& Meylan (1990), hereafter FM90;
Chmielewski et al. (1992), Neuforge-Verheecke \& Magain (1997),
Allende-Prieto et al (2004), hereafter ABLC04; del Peloso et al.
(2005a), Santos et al. (2005); Valenti \& Fisher (2005); Doyle et
al. (2005)), whilst five of them also performed this analysis for
the cooler and fainter component $\alpha$\,Cen\,B (Chmielewski et
al. 1992, Neuforge-Verheecke \& Magain (1997), ABLC04; Valenti \&
Fisher (2005) and Santos et al. 2005). All these authors, but
Chmielewski et al. (1992) and Santos et al. (2005), obtained
abundances for many chemical elements other than iron.

The analysis of FM90 for $\alpha$\,Cen\ A is noteworthy in that it
was the first to imply an abundance pattern considerably different
from solar, with excesses relative to Fe in Na, V, Mn, Co, Cu, and
deficits in Zn and the heavy neutron capture elements, and also
proposed a low \Teffb and a near solar metallicity for component
A, in contrast with most previously published figures. These
authors invoked a supernova to explain the peculiar chemical
features of the system. The next analysis (Chmielewski et al.
1992) sustained a high \Teffb and appreciably higher metallicity
for the system, which was also obtained by Neuforge-Verheecke \&
Magain (1997). The latter authors, moreover, found an abundance
pattern not diverging significantly from that of the Sun, though
supporting the deficiency of heavy elements found by FM90.

del Peloso et al. (2005a) and Santos et al. (2005) have both
derived a high metallicity for the system. Doyle et al. (2005)
added to the controversy by proposing both a low \Teffb and a
metallicity not appreciably above solar for component A, as did
FM90. Their abundance pattern is, however, solar. ABLC04 propose
for both components much lower \Teffsb than previously found by
any author. Even though their metallicity agrees reasonably with
that of Chmielewski et al. (1992) and Neuforge-Verheecke \& Magain
(1997), their detailed abundance pattern is highly non-solar and
also very different from any thus far, with high excesses of Mg,
Si, Ca, Sc, Ti, Zn and Y. Their low metallicity is a result of a
lower adopted \Teff, as also is the case for the FM90 analysis.

Doyle et al. (2005) presented the most recent abundance analysis
of $\alpha$\,Cen\,A and obtained abundances for six elements. They
made use of the Anstee, Barklem and O'Mara (ABO) line damping
theory (Barklem et al. 1998 and references therein), which allowed
them to fit accurate damping constants to the profile of strong
lines, turning these into reliable abundances indicators, an
approach normally avoided in abundance analyses. They found
[Fe/H]=$0.12\pm0.06$ for the iron abundance, which is in
disagreement with most authors using the standard method, although
agreeing with FM90. To bring home the point of the existing large
disagreement between the various published results, one needs look
no further than at the last entries of Table~\ref{hist-review},
all based on very high-quality data and state of the art methods.
These disagreements in chemical composition lie beyond the
confidence levels usually quoted by the authors. Moreover, the
dispersion of the \Teffb values found range between 300K and 400K,
respectively, for component A and B.

Pourbaix et al. (1999) finish their paper thus: ``we urge southern
spectroscopists to put a high priority on $\alpha$ Centauri''.
Clearly, this very important stellar system is entitled to
additional attention, fulfilling its utility as a reliable
calibrator for theories of stellar structure and evolution, and
taking full advantage of its tight observational constraints
towards our understanding of our second closest neighbor and the
atmospheres of cool stars. The widely differing results of the
chemical analysis also cast doubt about the place of $\alpha$
Centauri in the galactic chemical evolution scenario. The goal of
the present study is a simultaneous analysis of the two components
of the system, obtaining their atmospheric parameters and detailed
abundance pattern, providing an up-to-date comparative analysis of
the different determinations, the methods used and their results.

This paper is organized as follows. In section 2 we describe the
data acquisition and reduction. In section 3 we describe the
spectroscopic derivation of the atmospheric parameters and Fe
abundance, and compare them to other recent results from other
techniques, discussing possible sources of discrepancies. The
chemical composition pattern and its comparison to those of other
authors, is outlined in section 4. Section 5 is devoted to the
analysis of the evolutionary state of the system, and section 6
summarizes the conclusions.

\begin{figure}
\begin{center}
\resizebox{8.5cm}{!}{\includegraphics{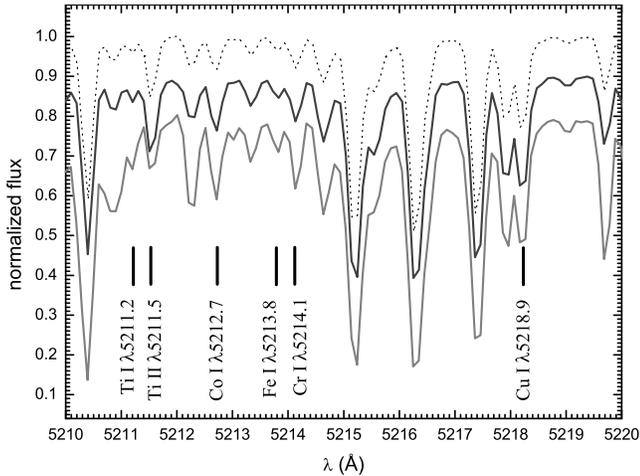}}
\end{center}
\caption[]{Sample spectra of the Moon, employed as a proxy for the
solar flux spectra, and $\alpha$ Centauri A and B. The nominal
resolution is R = 35\,000, and the signal-to-noise (S/N) ratio is
in excess of 1\,000. Some spectral lines measured for the
spectroscopic derivation of the atmospheric parameters and the
abundance analysis are indicated. The top, dotted spectrum is the
Moon's, the dark gray one corresponds to $\alpha$ Cen A, and the
light gray one to $\alpha$ Cen B. The $\alpha$ Cen spectra are
arbitrarily displaced vertically. The stronger line blocking in
$\alpha$ Cen B is apparent.} \label{sample-spectra}
\end{figure}

\section{Observations and line measurement}

Observations were performed in 2001 with the coud{\'e}
spectrograph, coupled to the 1.60m telescope of Observat{\'o}rio
do Pico dos Dias (OPD, Bras{\'o}polis, Brazil), operated by
Laborat{\'o}rio Nacional de Astrof{\i}sica (LNA/CNPq). As both
$\alpha$\,Cen\,A and B are solar-type stars, the Sun is the
natural choice as the standard star of a differential analysis.
The expectation of this approach is that systematic errors in the
measurement of line strengths, the representation of model
atmospheres, and the possible presence of Non-Local
Thermodynamic Equilibrium (NLTE) effects, will be eliminated or
at least greatly lessened, if the standard and the analysed object
are sufficiently similar. We chose the moon as a sunlight
surrogate to secure a solar flux spectrum. The slit width was
adjusted to give a two-pixel resolving power R = 35\,000. A 1\,800
l/mm diffraction grating was employed in the first direct order,
projecting onto a 24$\mu$m, 1024 pixels CCD. The exposure times
were chosen to allow for a S/N ratio in excess of 1\,000. A decker
was used to block one star of the binary system while exposing the
other, and we ascertained that there was no significant
contamination. The moon image, also exposed to very high S/N, was
stopped orthogonally to the slit width to a size comparable to the
seeing disks of the stars.

\begin{figure}
\begin{center}
\resizebox{8.5cm}{!}{\includegraphics{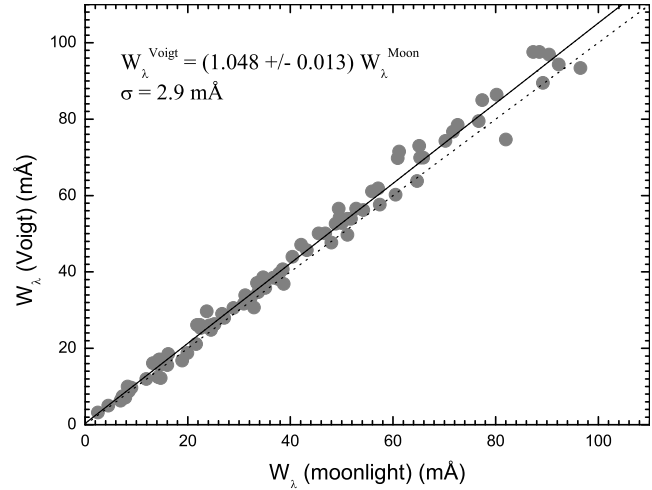}}
\end{center}
\caption[]{A plot of Voigt-fitted \Wlamsb from the Solar Flux
Atlas, given by Meylan et al. (1993), against our gaussian
measurements, for the 84 common lines. A linear regression defines
the correction to be applied to our measured \Wlamsb in order to
lessen systematic errors due to inadequate gaussian profile
fitting. As expected for non-saturated lines, we ascertained that
a linear relationship suffices to describe the correction. The
line wings, fitted incompletely by gaussian profiles, start to be
important for lines stronger than $\sim$50m{\AA}. The best fit
(solid line) and the one-to-one line of identity (dotted line) are
shown.} \label{gauss-voigt}
\end{figure}

Nine spectral regions were observed, centered at
$\lambda$$\lambda$ 5100, 5245, 5342, 5411, 5528, 5691, 5825, 6128
and 6242 {\AA}, with spectral coverage of 90\,{\AA} each. The
chemical species represented by spectral lines reasonably free
from blending are \ion{Na}{i}, \ion{Si}{i}, \ion{Ca}{i},
\ion{Sc}{i}, \ion{Sc}{ii}, \ion{Ti}{i}, \ion{Ti}{ii},\ion{V}{i},
\ion{Cr}{i}, \ion{Cr}{ii}, \ion{Mn}{i}, \ion{Fe}{i}, \ion{Fe}{ii},
\ion{Co}{i}, \ion{Ni}{i}, \ion{Cu}{i}, \ion{Y}{ii}, \ion{Ba}{ii}.
Additional data centered on the H$\alpha$ spectral region, for the
$\alpha$\,Cen stars and moonlight, were secured in 2004, using a
13.5$\mu$m, 4608 pixels CCD, integrated to S/N $\sim$ 500 and with
R = 43\,000.

Data reduction was carried out by the standard procedure using
IRAF{\footnote {{\it Image Reduction and Analysis Facility} (IRAF)
is distributed by the National Optical Astronomical Observatories
(NOAO), which is operated by the Association of Universities for
Research in Astronomy (AURA), Inc., under contract to the National
Science Foundation (NSF).}}. After usual bias and flat-field
correction, the background and scattered light were subtracted and
the one-dimensional spectra were extracted. No fringing was
present in our spectra. The pixel-to-wavelength calibration was
obtained from the stellar spectra themselves, by selecting
isolated spectral lines in the object spectra and checking for the
absence of blends, the main screen for blends being the Solar Flux
Atlas (Kurucz et al. 1984) and the Utrecht spectral line
compilation (Moore et al. 1966). Gaussian fits were applied to the
cores of the selected lines, and pixel-$\lambda$ polynomial fits
determined. For the short spectral selections individually
reduced, a 2nd-order polynomial always sufficed, the average
r.m.s. of the residuals being 0.005 ${\rm \AA}$ or better. There
followed the Doppler correction of all spectra to a rest reference
frame.

Normalization of the continuum is a very delicate and relevant
step in the analysis procedure, since the accuracy of line
equivalent width (hereafter \Wlam) measurements is very sensitive
to a faulty determination of the continuum level. We selected
continuum windows in the Solar Flux Atlas, apparently free from
telluric or photospheric lines. We took great care in constantly
comparing the spectra of the two $\alpha$\,Cen components and the
Sun, to ensure that a consistent choice of continuum windows was
achieved in all three objects, since the very strong-lined spectra
of the $\alpha$\,Cen\ stars caused continuum depressions
systematically larger than in the Sun. A number of pixels was
chosen in the selected continuum windows, followed by the
determination of a low order polynomial fitting these points. The
wavelength coverage of each single spectrum was in all cases
sufficient to ensure an appropriate number of windows, with
special attention given to the edge of the spectra. Sample spectra
are shown in Fig.~\ref{sample-spectra}. As will be seen below, the
errors of the atmospheric parameters derived directly from the
spectra, and the element abundances of $\alpha$\,Cen\,B, are
greater than in $\alpha$\,Cen\,A, probably due to a less
trouble-free normalization of its strongly line-blocked
spectrum, and to a better cancellation of uncertainties in
the differential analysis.

\begin{figure*}
\begin{center}
\resizebox{8.5cm}{!}{\includegraphics{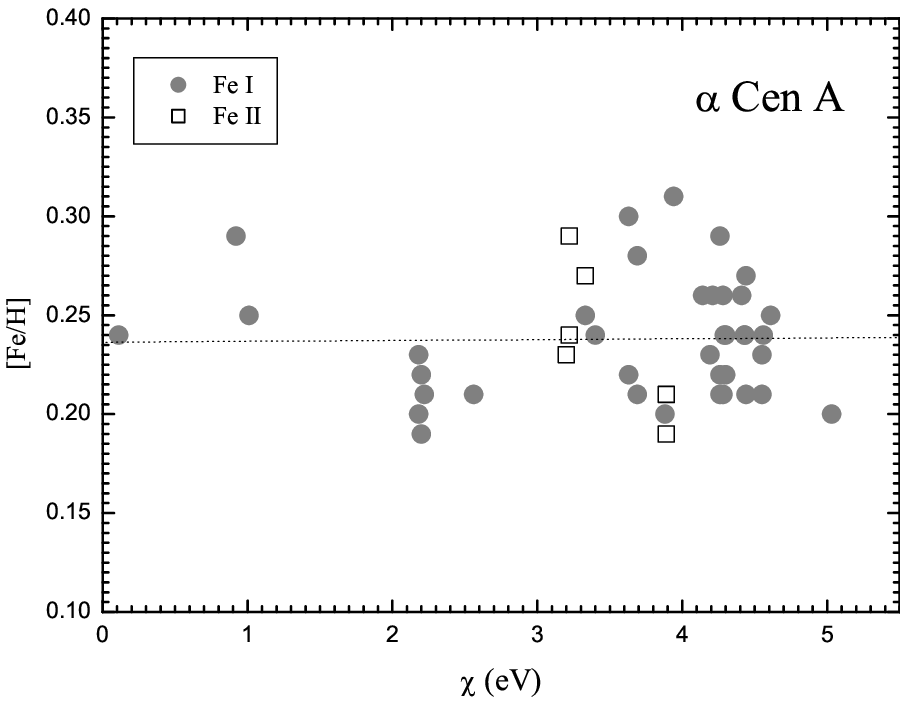}}
\resizebox{8.5cm}{!}{\includegraphics{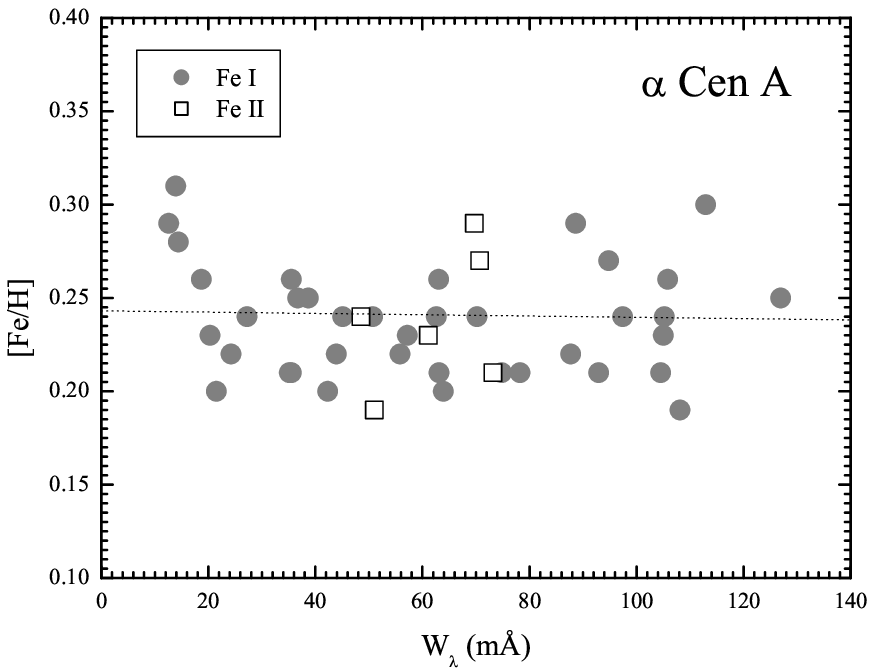}}
\end{center}
\caption[]{{\it Left}. \ion{Fe}{i} line abundances as a function
of excitation potential $\chi$ in $\alpha$\,Cen\,A. The dashed
line is the regression of the two quantities. By forcing the
angular coefficient to be null, we retrieve the excitation \Teffb of the star.

{\it Right}. \ion{Fe}{i} line abundances as a function of
equivalent width \Wlamb in $\alpha$\,Cen\,A. A null angular
coefficient provides the microturbulence velocity $\xi_t$. The
ionization equilibrium between \ion{Fe}{i} and \ion{Fe}{ii}
constrains the surface gravity, and is realized simultaneously
with the \Teffb and $\xi_t$ determinations. }
\label{method}
\end{figure*}

For the determination of element abundances, we chose lines of
moderate intensity with profiles that indicate little or no
blending. To avoid contamination of telluric lines we computed for
each spectrum, using cross-correlation techniques, the
displacement $\Delta\lambda$ the telluric lines would show
relative to their rest position $\lambda_0$, as given by the Solar
Flux Atlas. We discarded photospheric lines closer than
2$\Delta\lambda$ from a telluric line.

The equivalent widths were measured by fitting single or multiple
(the latter when deblending closely spaced lines) gaussian
profiles to the selected lines, using IRAF. The moderately high
spectral resolution we chose was designed to guarantee that the
instrumental profile dominates the observed profile, and therefore
that purely gaussian fits would adequately represent the observed
line profiles. To test the representation of the solar flux
spectrum by the moon, we also observed, with exactly the same
setup, spectra of daylight and the asteroid Ceres. A direct
comparison of the moon, daylight and asteroid \Wlamsb showed
perfect agreement between the three sets of measurements to better
than 1\% even for moderately strong lines. This lends confidence
to our determination of solar gf-values based on \Wlamb measured
off the moon spectra. The moonlight spectra was actually preferred
due to its higher S/N ratio as compared to Ceres, for which no
high-quality spectrum could be obtained in a reasonable exposure
time. Also, daylight spectra may show \Wlamb systematic fill-in
effects by up to 4\%, as a combination of aerosol and
Rayleigh-Brillouin effects (this effect depends on observing
angle and can be eliminated or minimized if care is applied, see 
Gray et al. 2000 for details). Even though no difference
could be measured in our spectra, we considered it more prudent to
use the moonlight spectrum as the solar proxy: it should be an
accurate representation of the solar flux spectrum in the visible.

Even at our not-so-high resolution, lines stronger than $\sim$50
m{\AA} begin to develop visible Voigt wings. To account for this
effect, we performed a linear regression of our gaussian moon
\Wlamsb against the \Wlamsb measured off the Solar Flux Atlas by
Meylan et al. (1993). These authors fitted Voigt profiles to a set
of lines deemed sufficiently unblended to warrant the measurement
of their true \Wlams, and they should be a homogeneous and
high-precision representation of the true line intensities. We
then determined the correction necessary to convert our
measurements to a scale compatible with the Voigt-fitted \Wlams.
The result is shown in Fig.~\ref{gauss-voigt}, where the excellent
correlation, with very small dispersion, is seen. The correction
derived is

\begin{equation}
{
 W_\lambda^{\rm Voigt} = (1.048 \pm 0.013) W_\lambda^{\rm moon}
}
\end{equation}

The {\it r.m.s.} standard deviation of the linear regression is
2.9 m{\AA}, regarding the Voigt \Wlamb as essentially error-free
as compared to our data. This regression was applied to all our
\Wlamb measurements. We take 2.9 m{\AA} as the 1$\sigma$
uncertainty of our internal \Wlamb measurements.

\begin{table*}

\caption[]{Spectral lines of the elements used in this study. A 
and B refer to $\alpha$\,Cen\,A and $\alpha$\,Cen\,B. \Wlamsb are 
the raw measurements, prior to the correction to the Voigt \Wlamb 
scale (see text).}

\label{line-data}

\begin{center}

\begin{tabular}{l c c c c c   c   l c c c c c   c   l c c c c c}\hline

\s$\lambda$&\s$\chi$&\s$\log gf$&\multicolumn{3}{c}{\s\Wlamb
(m\AA)} &\p&

\s$\lambda$&\s$\chi$&\s$\log gf$&\multicolumn{3}{c}{\s\Wlamb
(m\AA)} &\p&

\s$\lambda$&\s$\chi$&\s$\log gf$&\multicolumn{3}{c}{\s\Wlamb
(m\AA)}     \\\cline{4-6}\cline{11-13}\cline{18-20}

\s(\AA)    &\s(eV)  &\p         &\p Moon &\p A &\p B &\p&

\s(\AA)    &\s(eV)  &\p         &\p Moon &\p A &\p B &\p&

\s(\AA)    &\s(eV)  &\p         &\p Moon &\p A &\p B \\\hline

\multicolumn{6}{c}{\p Na I}                       &\p&\p5381.020&\p1.57&\p-1.855&\p 65.9&\p 79.0&\p 80.9&\p&\p5701.557&\p2.56&\p-2.116&\p 89.2&\p 99.5&\p127.9  \\

\p6154.230&\p2.10&\p-1.532&\p 40.6&\p 62.5&\p 92.3&\p&\p5418.756&\p1.58&\p-2.116&\p 52.8&\p 66.9&\p 55.8&\p&\p5705.473&\p4.30&\p-1.427&\p 42.1&\p 53.0&\p 66.8  \\

\p6160.753&\p2.10&\p-1.224&\p 60.5&\p 86.3&\p120.4&\p&\multicolumn{6}{c}{\p}                &\p&\p5731.761&\p4.26&\p-1.115&\p 60.9&\p 71.1&\p 85.9  \\

\p        &\p    &\p      &\p     &\p     &\p     &\p&\multicolumn{6}{c}{\p V I}                        &\p&\p5784.666&\p3.40&\p-2.487&\p 33.0&\p 42.8&\p 61.0  \\

\multicolumn{6}{c}{\p Mg I}                       &\p&\p5657.436&\p1.06&\p-0.889&\p  8.5&\p 10.4&\p 40.4&\p&\p5811.916&\p4.14&\p-2.383&\p 11.3&\p 17.6&\p 26.3  \\

\p5711.095&\p4.34&\p-1.658&\p112.2&\p124.3&\p  ...&\p&\p5668.362&\p1.08&\p-0.940&\p  7.3&\p 11.7&\p 38.2&\p&\p5814.805&\p4.28&\p-1.851&\p 23.7&\p 33.6&\p 44.7  \\

\p5785.285&\p5.11&\p-1.826&\p 59.0&\p 68.3&\p 88.7&\p&\p5670.851&\p1.08&\p-0.396&\p 21.6&\p 31.4&\p 82.1&\p&\p5835.098&\p4.26&\p-2.085&\p 16.2&\p 22.8&\p 35.0  \\

\p        &\p    &\p      &\p     &\p     &\p     &\p&\p5727.661&\p1.05&\p-0.835&\p  9.8&\p 14.6&\p 56.8&\p&\p5849.681&\p3.69&\p-2.963&\p  8.3&\p 13.5&\p 22.9  \\

\multicolumn{6}{c}{\p Si I}                       &\p&\p6135.370&\p1.05&\p-0.674&\p 14.1&\p 20.6&\p 59.6&\p&\p5852.222&\p4.55&\p-1.180&\p 43.2&\p 54.3&\p 71.3  \\

\p5517.533&\p5.08&\p-2.454&\p 14.5&\p 25.5&\p 21.8&\p&\p6150.154&\p0.30&\p-1.478&\p 12.9&\p 20.4&\p 65.5&\p&\p5855.086&\p4.61&\p-1.521&\p 24.5&\p 34.8&\p 42.8  \\

\p5665.563&\p4.92&\p-1.957&\p 43.0&\p 57.0&\p 62.3&\p&\p6274.658&\p0.27&\p-1.570&\p 11.1&\p 15.0&\p 60.6&\p&\p5856.096&\p4.29&\p-1.553&\p 36.6&\p 48.1&\p 61.2  \\

\p5684.484&\p4.95&\p-1.581&\p 65.1&\p 79.4&\p 78.6&\p&\p6285.165&\p0.28&\p-1.543&\p 11.5&\p 25.1&\p 66.3&\p&\p5859.596&\p4.55&\p-0.579&\p 77.4&\p 88.4&\p105.6  \\

\p5690.433&\p4.93&\p-1.627&\p 63.2&\p 67.4&\p 64.6&\p&\p        &\p    &\p      &\p     &\p     &\p     &\p&\p6098.250&\p4.56&\p-1.760&\p 17.7&\p 25.7&\p 35.4  \\

\p5701.108&\p4.93&\p-1.967&\p 41.9&\p 57.3&\p 58.2&\p&\multicolumn{6}{c}{\p Cr I}                       &\p&\p6120.249&\p0.92&\p-5.733&\p  7.5&\p 11.8&\p 29.1  \\

\p5708.405&\p4.95&\p-1.326&\p 78.7&\p 93.1&\p 95.9&\p&\p5214.144&\p3.37&\p-0.739&\p 18.2&\p 27.2&\p 39.9&\p&\p6137.002&\p2.20&\p-2.830&\p 72.9&\p 83.4&\p104.9  \\

\p5793.080&\p4.93&\p-1.896&\p 46.1&\p 62.4&\p 60.3&\p&\p5238.964&\p2.71&\p-1.312&\p 19.9&\p 33.8&\p 55.2&\p&\p6151.616&\p2.18&\p-3.308&\p 51.1&\p 60.7&\p 80.5  \\

\p6125.021&\p5.61&\p-1.496&\p 34.7&\p 51.3&\p 51.4&\p&\p5272.007&\p3.45&\p-0.311&\p 32.9&\p 42.9&\p 88.0&\p&\p6173.340&\p2.22&\p-2.871&\p 70.2&\p 83.7&\p100.9  \\

\p6142.494&\p5.62&\p-1.422&\p 38.5&\p 52.5&\p 49.8&\p&\p5287.183&\p3.44&\p-0.822&\p 13.8&\p 16.6&\p 34.9&\p&\p6219.287&\p2.20&\p-2.412&\p 93.6&\p102.9&\p134.6  \\

\p6145.020&\p5.61&\p-1.397&\p 40.5&\p 55.1&\p 55.3&\p&\p5296.691&\p0.98&\p-1.343&\p 96.5&\p107.0&\p153.1&\p&\p6226.730&\p3.88&\p-2.068&\p 31.2&\p 40.1&\p 56.8  \\

\p6243.823&\p5.61&\p-1.220&\p 51.8&\p 67.5&\p 64.1&\p&\p5300.751&\p0.98&\p-2.020&\p 64.7&\p 76.6&\p103.7&\p&\p6240.645&\p2.22&\p-3.295&\p 50.0&\p 60.0&\p 80.4  \\

\p6244.476&\p5.61&\p-1.264&\p 48.8&\p 66.5&\p 66.3&\p&\p5304.183&\p3.46&\p-0.701&\p 16.8&\p 25.1&\p 16.8&\p&\p6265.131&\p2.18&\p-2.537&\p 88.3&\p 99.9&\p135.0  \\

\p        &\p    &\p      &\p     &\p     &\p     &\p&\p5318.810&\p3.44&\p-0.647&\p 19.2&\p 27.9&\p 53.4&\p&\p6271.283&\p3.33&\p-2.703&\p 26.5&\p 36.7&\p 54.9  \\

\multicolumn{6}{c}{\p Ca I}                       &\p&\p5784.976&\p3.32&\p-0.360&\p 34.0&\p 48.0&\p 72.9&\p&\p        &\p    &\p      &\p     &\p     &\p       \\

\p5261.708&\p2.52&\p-0.564&\p126.9&\p123.0&\p182.1&\p&\p5787.965&\p3.32&\p-0.129&\p 49.4&\p 59.8&\p 83.8&\p&\multicolumn{6}{c}{\p Fe II }                       \\

\p5867.572&\p2.93&\p-1.566&\p 26.7&\p 35.8&\p 58.6&\p&\p        &\p    &\p      &\p     &\p     &\p     &\p&\p5234.630&\p3.22&\p-2.199&\p 90.4&\p110.9&\p 92.7  \\

\p6161.295&\p2.52&\p-1.131&\p135.5&\p 85.7&\p128.7&\p&\multicolumn{6}{c}{\p Cr II}                      &\p&\p5264.812&\p3.33&\p-2.930&\p 53.1&\p 67.2&\p 64.8  \\

\p6163.754&\p2.52&\p-1.079&\p126.6&\p 87.7&\p 93.8&\p&\p5305.855&\p3.83&\p-2.042&\p 27.1&\p 37.8&\p 24.1&\p&\p5325.560&\p3.22&\p-3.082&\p 51.1&\p 66.3&\p 52.7  \\

\p6166.440&\p2.52&\p-1.116&\p 72.6&\p 88.9&\p114.0&\p&\p5313.526&\p4.07&\p-1.539&\p 38.7&\p 49.3&\p 43.4&\p&\p5414.075&\p3.22&\p-3.485&\p 33.7&\p 46.0&\p 32.1  \\

\p6169.044&\p2.52&\p-0.718&\p 97.7&\p113.4&\p158.0&\p&\p        &\p    &\p      &\p     &\p     &\p     &\p&\p5425.257&\p3.20&\p-3.229&\p 45.5&\p 58.1&\p 46.3  \\

\p6169.564&\p2.52&\p-0.448&\p118.5&\p137.5&\p185.8&\p&\multicolumn{6}{c}{\p Mn I}                       &\p&\p6149.249&\p3.89&\p-2.761&\p 37.8&\p 48.4&\p 30.0  \\

\p        &\p    &\p      &\p     &\p     &\p     &\p&\p5394.670&\p0.00&\p-2.916&\p 83.4&\p103.8&\p165.0&\p&\p6247.562&\p3.89&\p-2.325&\p 57.1&\p 69.6&\p 48.3  \\

\multicolumn{6}{c}{\p  Sc I}                      &\p&\p5399.479&\p3.85&\p-0.045&\p 42.9&\p 63.4&\p 96.6&\p&\multicolumn{6}{c}{\p }                             \\

\p5671.826&\p1.45&\p 0.538&\p 16.1&\p 24.2&\p 64.1&\p&\p5413.684&\p3.86&\p-0.343&\p 28.0&\p 45.8&\p 73.1&\p&\multicolumn{6}{c}{\p Co I  }                       \\

\p6239.408&\p0.00&\p-1.270&\p  7.8&\p 12.3&\p 12.9&\p&\p5420.350&\p2.14&\p-0.720&\p 88.6&\p116.4&\p177.5&\p&\p5212.691&\p3.51&\p-0.180&\p 20.0&\p 34.8&\p 50.6  \\

\p        &\p    &\p      &\p     &\p     &\p     &\p&\p5432.548&\p0.00&\p-3.540&\p 54.9&\p 73.3&\p141.0&\p&\p5301.047&\p1.71&\p-1.864&\p 22.5&\p 33.8&\p 57.9  \\

\multicolumn{6}{c}{\p Sc II}                      &\p&\p5537.765&\p2.19&\p-1.748&\p 37.3&\p 57.4&\p113.3&\p&\p5342.708&\p4.02&\p 0.661&\p 35.1&\p 48.5&\p 80.5  \\

\p5318.346&\p1.36&\p-1.712&\p 16.0&\p 23.3&\p 28.7&\p&\p        &\p    &\p      &\p     &\p     &\p     &\p&\p5359.192&\p4.15&\p 0.147&\p 11.9&\p 20.1&\p 37.3  \\

\p5526.815&\p1.77&\p 0.099&\p 80.2&\p 97.8&\p 86.5&\p&\multicolumn{6}{c}{\p Fe I}                       &\p&\p5381.772&\p4.24&\p 0.000&\p  8.2&\p 15.6&\p 20.0  \\

\p5657.874&\p1.51&\p-0.353&\p 71.2&\p 85.9&\p 76.5&\p&\p5054.647&\p3.64&\p-1.806&\p 53.0&\p 67.6&\p 97.6&\p&\p5454.572&\p4.07&\p 0.319&\p 18.9&\p 26.8&\p 42.8  \\

\p5684.189&\p1.51&\p-0.984&\p 40.4&\p 55.6&\p 48.9&\p&\p5067.162&\p4.22&\p-0.709&\p 85.4&\p 90.3&\p123.1&\p&\multicolumn{6}{c}{\p}                              \\

\p6245.660&\p1.51&\p-1.063&\p 37.6&\p 51.1&\p 51.0&\p&\p5109.649&\p4.30&\p-0.609&\p 87.9&\p100.1&\p143.0&\p&\multicolumn{6}{c}{\p Ni I}                         \\

\p        &\p    &\p      &\p     &\p     &\p     &\p&\p5127.359&\p0.93&\p-3.186&\p109.6&\p117.7&\p164.8&\p&\p5094.406&\p3.83&\p-1.088&\p 32.4&\p 45.0&\p 62.2  \\

\multicolumn{6}{c}{\p Ti I}                       &\p&\p5151.971&\p1.01&\p-3.128&\p108.9&\p120.8&\p177.5&\p&\p5220.300&\p3.74&\p-1.263&\p 28.5&\p 40.6&\p 52.1  \\

\p5071.472&\p1.46&\p-0.683&\p 36.1&\p 45.7&\p 99.0&\p&\p5213.818&\p3.94&\p-2.752&\p  7.5&\p 13.0&\p 20.2&\p&\p5435.866&\p1.99&\p-2.340&\p 57.4&\p 73.8&\p 88.9  \\

\p5113.448&\p1.44&\p-0.815&\p 30.9&\p 36.4&\p 98.3&\p&\p5223.188&\p3.63&\p-2.244&\p 32.3&\p 41.6&\p 56.5&\p&\p5452.860&\p3.84&\p-1.420&\p 19.0&\p 29.1&\p 37.3  \\

\p5145.464&\p1.46&\p-0.615&\p 39.6&\p 50.8&\p 90.6&\p&\p5225.525&\p0.11&\p-4.577&\p 83.2&\p 92.7&\p128.8&\p&\p5846.986&\p1.68&\p-3.380&\p 24.2&\p 35.8&\p 56.9  \\

\p5147.479&\p0.00&\p-1.973&\p 43.7&\p 66.8&\p102.7&\p&\p5242.491&\p3.63&\p-1.083&\p 92.3&\p107.5&\p132.2&\p&\p6176.807&\p4.09&\p-0.315&\p 61.2&\p 86.8&\p 90.1  \\

\p5152.185&\p0.02&\p-2.130&\p 34.9&\p 43.5&\p 79.1&\p&\p5243.773&\p4.26&\p-0.947&\p 69.6&\p 84.3&\p 91.5&\p&\p6177.236&\p1.83&\p-3.476&\p 16.4&\p 29.4&\p 41.2  \\

\p5211.206&\p0.84&\p-2.063&\p  9.3&\p 13.7&\p  ...&\p&\p5250.216&\p0.12&\p-4.668&\p 78.1&\p 96.1&\p136.0&\p&\p        &\p    &\p      &\p     &\p     &\p       \\

\p5219.700&\p0.02&\p-2.264&\p 28.9&\p 39.0&\p 82.9&\p&\p5321.109&\p4.43&\p-1.191&\p 48.0&\p 59.5&\p 76.8&\p&\multicolumn{6}{c}{\p Cu I  }                       \\

\p5295.780&\p1.07&\p-1.633&\p 14.2&\p 18.6&\p 49.8&\p&\p5332.908&\p1.56&\p-2.751&\p102.8&\p118.6&\p139.8&\p&\p5218.209&\p3.82&\p 0.293&\p 55.9&\p 71.8&\p 82.5  \\

\p5426.236&\p0.02&\p-2.903&\p  9.0&\p 12.1&\p 48.2&\p&\p5379.574&\p3.69&\p-1.542&\p 64.8&\p 76.5&\p 94.4&\p&\p5220.086&\p3.82&\p-0.630&\p 15.5&\p 25.0&\p 32.5  \\

\p5679.937&\p2.47&\p-0.535&\p  8.6&\p 10.1&\p 28.8&\p&\p5389.486&\p4.41&\p-0.533&\p 87.3&\p102.2&\p123.0&\p&\p        &\p    &\p      &\p     &\p     &\p       \\

\p5866.452&\p1.07&\p-0.842&\p 49.6&\p 64.8&\p107.5&\p&\p5395.222&\p4.44&\p-1.653&\p 25.2&\p 33.7&\p 52.1&\p&\multicolumn{6}{c}{\p Y II}                         \\

\p6098.694&\p3.06&\p-0.095&\p  6.9&\p 10.1&\p 27.6&\p&\p5432.946&\p4.44&\p-0.682&\p 76.4&\p 90.2&\p106.7&\p&\p5087.426&\p1.08&\p-0.329&\p 49.3&\p 53.7&\p 54.2  \\

\p6126.224&\p1.07&\p-1.358&\p 25.3&\p 31.9&\p 68.3&\p&\p5491.845&\p4.19&\p-2.209&\p 14.3&\p 25.4&\p 32.9&\p&\p5289.820&\p1.03&\p-1.847&\p  4.5&\p  5.6&\p  7.7  \\

\p6258.104&\p1.44&\p-0.410&\p 54.2&\p 65.5&\p102.5&\p&\p5522.454&\p4.21&\p-1.418&\p 46.8&\p 59.9&\p 70.4&\p&\p5402.780&\p1.84&\p-0.510&\p 14.7&\p 22.2&\p 22.1  \\

\p        &\p    &\p      &\p     &\p     &\p     &\p&\p5560.207&\p4.43&\p-1.064&\p 55.1&\p 66.7&\p 75.6&\p&\p        &\p    &\p      &\p     &\p     &\p       \\

\multicolumn{6}{c}{\p Ti II}                      &\p&\p5577.013&\p5.03&\p-1.415&\p 14.5&\p 20.3&\p 26.3&\p&\multicolumn{6}{c}{\p Ba II}                        \\

\p5211.544&\p2.59&\p-1.551&\p 33.5&\p 46.1&\p 35.2&\p&\p5661.348&\p4.28&\p-1.802&\p 25.5&\p 33.6&\p 48.6&\p&\p5853.688&\p0.60&\p-0.828&\p 67.5&\p 73.4&\p 68.6  \\

\p5336.783&\p1.58&\p-1.592&\p 76.7&\p 90.0&\p
89.7&\p&\p5680.240&\p4.19&\p-2.255&\p 13.2&\p 19.1&\p
27.8&\p&\p6141.727&\p0.70&\p 0.244&\p124.4&\p127.4&\p140.1
\\\hline

\end{tabular}

\end{center}

\end{table*}

\section{Atmospheric parameters and Fe abundance}

A solar gf-value for each spectral line was calculated from a LTE,
1-D, homogeneous and plane-parallel solar model atmosphere from
the NMARCS grid, as described by Edvardsson et al. (1993) (see
http://marcs.astro.uu.se, and also Gustafsson et al. 2008). The
adopted parameters for the Sun were \Teff = 5780K, log g = 4.44,
$[$Fe/H$]$ = $+$0.00 and $\xi_t$ = 1.30 km\,s$^{\rm -1}$, and we
employed the \Wlamsb measured off the moon spectra, corrected to
the Voigt scale. The adopted solar absolute abundances are those
of Grevesse \& Noels (1993). In a purely differential analysis
such as ours, the absolute abundance scale is inconsequential. We
provide in Table~\ref{line-data} the details of all lines used.
They include wavelength $\lambda$, excitation potential $\chi$,
the calculated solar log $gf$ values, and the raw measured \Wlamsb
in the moon's, $\alpha$\,Cen\,A and $\alpha$\,Cen\,B spectra,
prior to the correction to the Voigt system
(Fig.~\ref{gauss-voigt}). Hyperfine structure (HFS) corrections
for the lines of \ion{Mg}{i}, \ion{Sc}{i}, \ion{Sc}{ii},
\ion{V}{i}, \ion{Mn}{i}, \ion{Co}{i}, \ion{Cu}{i} and \ion{Ba}{ii}
were adopted from Steffen (1985). del Peloso et al. (2005b)
discuss the influence of adopting different HFS scales on
abundance analyses of Mn and Co, concluding that it is
small, particularly for metallicities not too far from the solar
one, as compared to not using any HFS data. Therefore, the source
of the HFS corrections is not an important issue on the error
budget of our analysis, at least for Mn and Co. The other
elements of our analysis requiring HFS have usually simpler
structures (excepting Cu), and it is safe to conclude that the use
of HFS has not introduced any important error.

The atmospheric parameters of the $\alpha$\,Cen\ stars were
determined by simultaneously realizing the excitation \&
ionization equilibria of \ion{Fe}{i} and \ion{Fe}{ii}. For each
star, \Teffb was obtained by forcing the \ion{Fe}{i} line
abundances to be independent of their excitation potential.
Surface gravity was determined by forcing the lines of \ion{Fe}{i}
and \ion{Fe}{ii} to yield the same abundance. Microturbulence
velocities $\xi_t$ were determined forcing the lines of
\ion{Fe}{i} to be independent of their \Wlams. The Fe abundance
$[$Fe/H$]$ (we use throughout the notation $[$A/B$]$ = log
N(A)/N(B)$_{\rm star}$ - log N(A)/N(B)$_{\rm Sun}$, where N
denotes the number abundance) is automatically obtained as a
byproduct of this method. The solution thus obtained is unique for
a given set of gf values, \Wlamsb and model atmospheres, being
independent of the starting point and the iteration path. The
spectrum synthesis code is originally due to Dr. Monique Spite
(Observatoire de Paris-Meudon, Spite 1967), having been
continuously up-dated in the last 40 years.

Formal errors are estimated as follows: for \Teff, the 1$\sigma$
uncertainty of the slope of the linear regression in the [Fe/H]
vs. $\chi$ diagram yields the \Teffb variation which could still
be accepted at the 1$\sigma$ level. For the microturbulence
velocity, the same procedure provides the 1$\sigma$
microturbulence uncertainty in the [Fe/H] vs. \Wlamb diagram. For
the metallicity [Fe/H], we adopt the standard deviation of the
distribution of abundances derived from the \ion{Fe}{i} lines. The
error in $\log g$ is estimated by evaluating the variation in this
parameter which produces a disagreement of 1$\sigma$ between the
abundances of \ion{Fe}{i} and \ion{Fe}{ii}, where we regarded
the abundance offset as 1$\sigma$ when its value was equal to the
largest dispersion of the Fe abundances (usually that of
\ion{Fe}{ii}). The results of this procedure are shown in
Fig.~\ref{method}, where we plot the iron abundances of
$\alpha$\,Cen\,A derived from lines of both \ion{Fe}{i} and
\ion{Fe}{ii} against the excitation potential and \Wlams. The
baseline of the \ion{Fe}{i} lines is seen to be large both in
\Wlamb and $\chi$.

\begin{figure*}
\begin{center}
\resizebox{16cm}{!}{\includegraphics[angle=270]{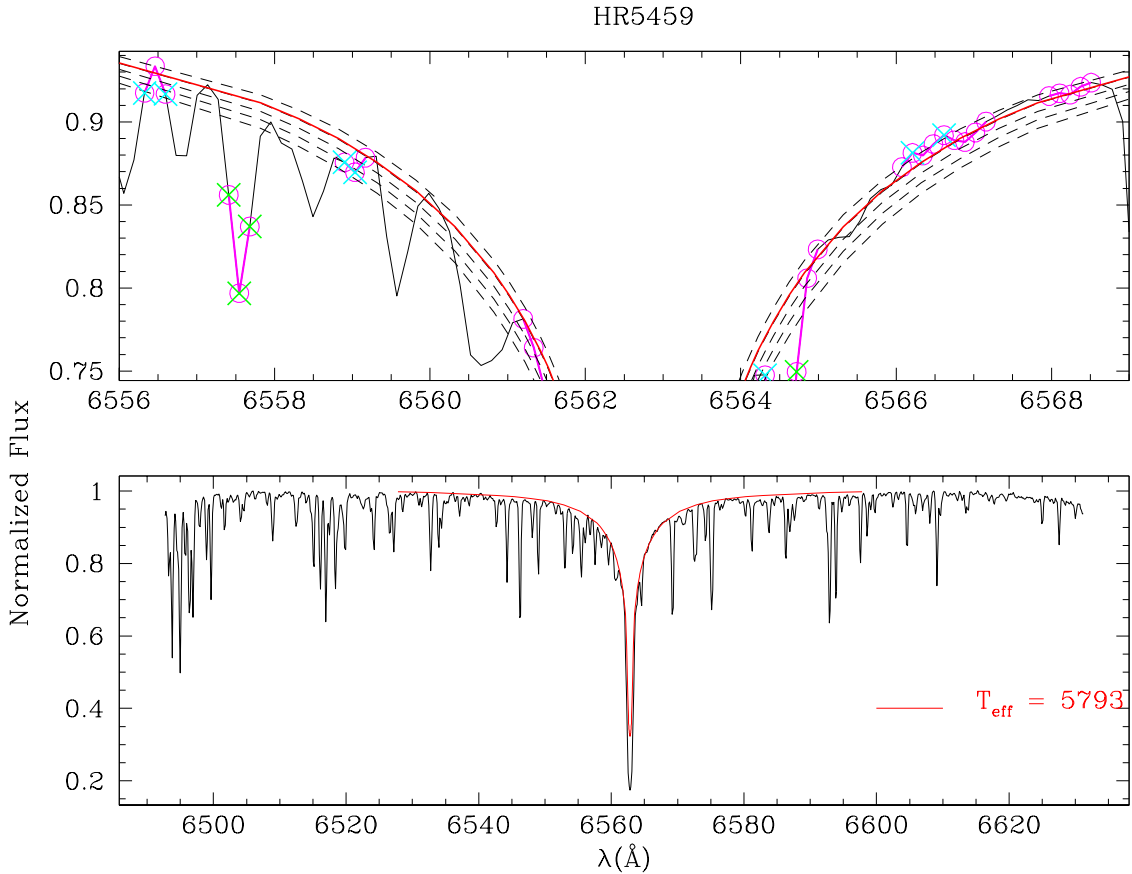}}
\end{center}
\caption[]{Effective temperature determination by fitting
theoretical profiles to the wings of H$\alpha$ for
$\alpha$\,Cen\,A. Upper panel: crosses refer to pixels
eliminated by the statistical test and 2$\sigma$ criterion (see
Lyra \& Porto de Mello 2005 for details). The \Teffb derived by the
accepted pixels (open circles) is 5793\,K. The corresponding 
best determined line profile is overplotted (solid thick line) on
the observed spectrum. The dashed lines are theoretical profiles,
spaced by 50 K and centered in 5847\,K, the spectroscopic \Teff.
Lower panel: the spectrum at a larger scale.}
\label{alfa-alfa}
\end{figure*}

\begin{figure*}
\begin{center}
\resizebox{16cm}{!}{\includegraphics[angle=270]{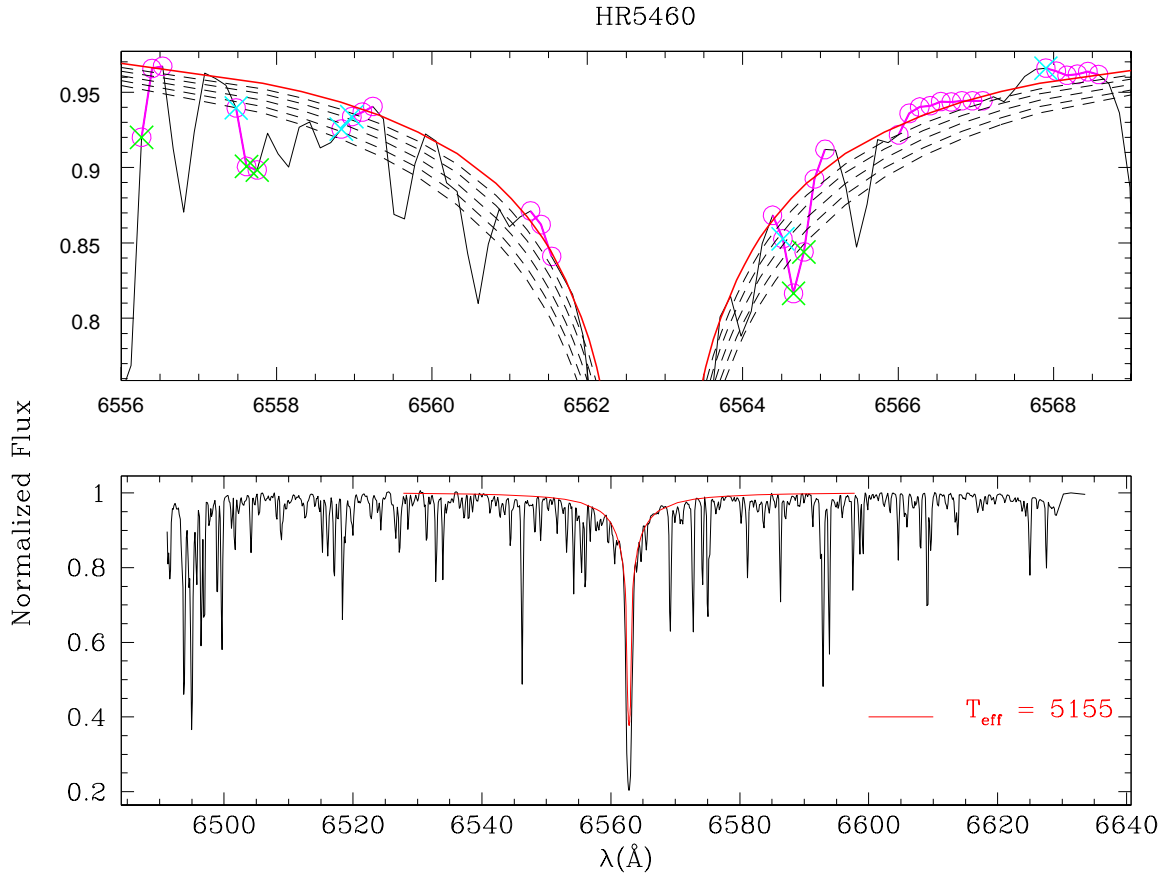}}
\end{center}
\caption[]{The same as in Fig~\ref{alfa-alfa} for
$\alpha$\,Cen\,B. The \Teffb derived by the accepted pixels is
5155\,K. The corresponding line profile is overplotted (solid
thick line) on the observed spectrum. The dashed lines are
theoretical profiles, spaced by 50 K and centered in 5316\,K, the
spectroscopic \Teff. Lower panel: the spectrum at a larger
scale.} \label{beta-alfa}
\end{figure*}

Additional effective temperatures were determined by fitting the
observed wings of H$\alpha$, using the automated procedure
described in detail in Lyra \& Porto de Mello (2005). We have
employed for the $\alpha$\,Cen\ stars new spectroscopic data from
the Observat\'orio do Pico dos Dias, with the same resolution but
greater signal-to-noise ratio than used by Lyra \& Porto de Mello
(2005). This procedure is shown in Fig.~\ref{alfa-alfa} and
Fig.~\ref{beta-alfa}. We found \Teff = 5793 $\pm$ 25 K
($\alpha$\,Cen\,A) and \Teff = 5155 $\pm$ 4 K ($\alpha$\,Cen\,B).
The moon spectrum is very well fitted by the parameters adopted
for the NMARCS solar atmosphere model. The quoted standard errors
refer exclusively to the dispersion of temperature values
attributed to the fitted profile data points. This makes the
uncertainty of the $\alpha$\,Cen\,B \Teffb artificially very low,
due to the high number of rejected points. An analysis of errors
incurred by the atmospheric parameters assumed in the fitting
procedure, plus the photon statistics (not including possible
systematic effects produced by the modelling, see Lyra \& Porto de
Mello (2005) for a full discussion), points to an average error of
$\sim$50 K in the \Teffsb determined from the H$\alpha$ line. For
the very low-noise spectra of these two stars, the expected errors
would be slightly less, but the greater difficulty in finding
line-free sections in the H$\alpha$ profile of the severely
blended spectrum of $\alpha$\,Cen\,B offsets this advantage. For
the latter, thus, the probable uncertainty should be closer to
$\sim$100 K. The normalization procedure is a a relevant source of
error for \Teffsb derived from H$\alpha$, as discussed by Lyra \&
Porto de Mello (2005): they found that a 0.2\% error in the
continuum level translates to $\sim$25 K in \Teff.

\begin{figure*}
\begin{center}
\resizebox{16cm}{!}{\includegraphics[angle=0]{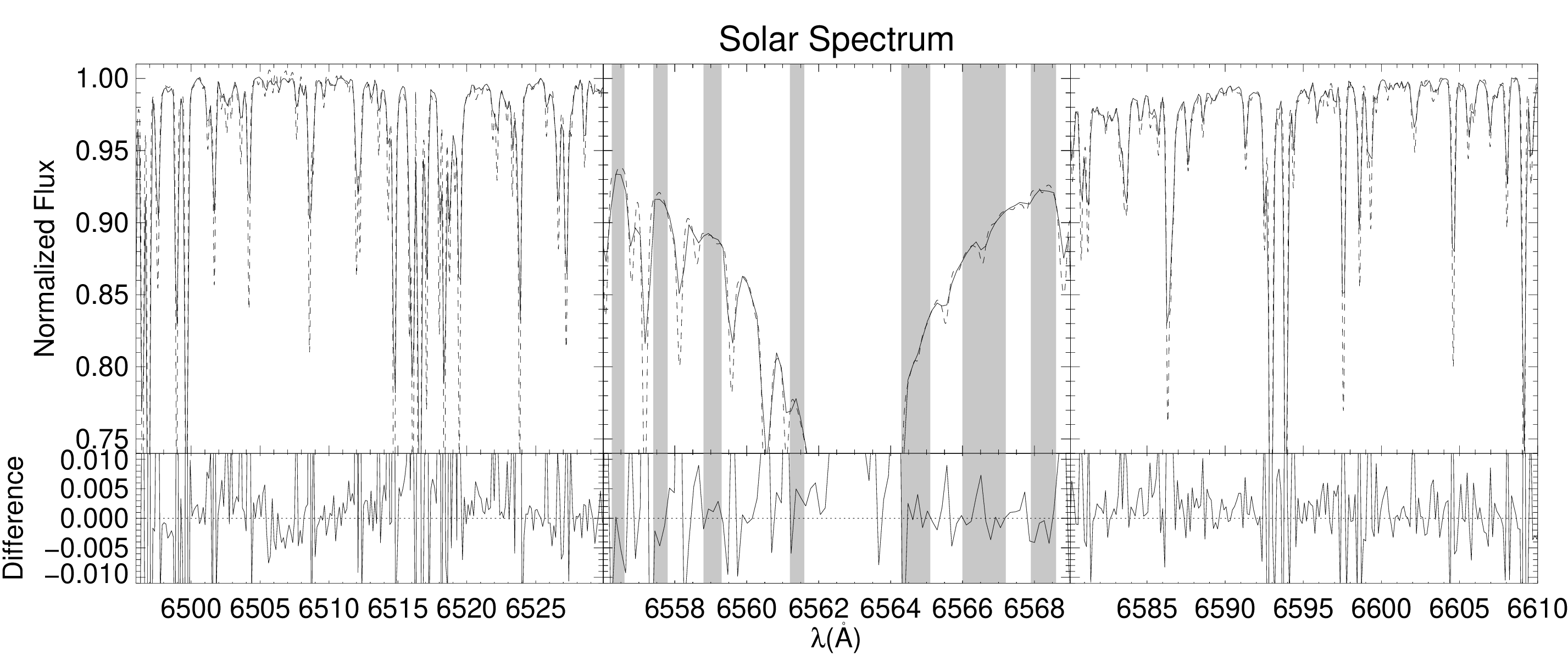}}
\end{center}
\caption[]{A comparison of the normalization of our H$\alpha$
spectrum of the Sun (moon) with that of ABLC04. The large
differences in the telluric lines are apparent. The left and right
panels depict the regions outside the H$\alpha$ profile which were
used for the continuum normalization. In the center panel we
depict, as the shaded gray areas, the wavelengths of the wing
profile used for the \Teffb determination. Note that the abscissa
is truncated to show only the most relevant portions of the
spectra, the blue and red limits, which set the normalization
scale, and those used directly for the fit of the line wings. In
this example, the mean difference between the two spectra
(computed only for the line wing regions actually fitted) is -0.8
$\pm$ 0.7 \%, and it is the worst case of our comparison of eight
spectra (see text for details).} \label{allende}
\end{figure*}

As an external check on our normalization procedure, we have
compared H$\alpha$ spectra of the Sun (moon) from Lyra \& Porto de
Mello (2005), from eight independent observing runs, also reduced
independently, with the data of ABLC04, who performed a
careful two-dimensional continuum normalization of the echelle
spectra. Due to normalization problems in the FEROS/ESO (La
Silla) spectra, these authors employed H$\alpha$ spectra as a
\Teffb criterion only for the northern stars of their sample, which
could be observed with the McDonald 2dcoud\'e spectrograph. So a
direct comparison between the two sets of H$\alpha$ spectra is not
possible for the $\alpha$\,Cen stars. We found, for the solar
spectra, an average difference of only ($-$0.29 $\pm$ 0.39) \%, as
measured in those regions relevant to the \Teffb determination.
This assures us of the absence of important systematic errors in
this respect. A comparison of the H$\alpha$ normalized spectra is
shown in Fig.~\ref{allende}.

Another check on the \Teffb values of the two stars may be
obtained from the IRFM scale of Ram\'{\i}rez \& Mel\'endez
(2005a). Ram\'{\i}rez \& Mel\'endez (2005b) compare, for the two
$\alpha$\,Cen\ stars, direct \Teffs, obtained from measured
bolometric fluxes and angular diameters, with those determined
from the IRFM method, as well those obtained from the application
of their own \Teff(IRFM) scale to color indices and an adopted
metallicity of [Fe/H] = $+$0.20, for both stars. Respectively,
they find, for $\alpha$\,Cen\,A, \Teff (direct) = 5771, \Teff
(IRFM) = 5759 K and \Teff (calibration) = 5736 K; the
corresponding values for $\alpha$\,Cen\,B are, respectively, \Teff
(direct) = 5178 K, \Teff (IRFM) = 5221 K and \Teff (calibration) =
5103 K. They adopt as weighted averages \Teff ($\alpha$\,Cen\,A) =
5744 $\pm$ 72 K and \Teff ($\alpha$\,Cen\,B) = 5136 $\pm$ 68 K.
For $\alpha$\,Cen\,A, a good agreement is found between this \Teffb
and our H$\alpha$ one. Formally, there is also a reasonable
agreement between this \Teffb and our spectroscopic one. For
$\alpha$\,Cen\,B, however, the spectroscopic \Teffb is
significantly higher than those derived from H$\alpha$ and the
IRFM method.

Y{\i}ld{\i}z (2007) drew attention to often neglected BVRI (Cousins
system) measurements of the system's components by Bessell (1990).
It has long been considered risky to use color indices for the
\Teffb determination of very bright stars, in which a variety of
systematic effects are expected as compared to standard stars of
photometric systems, among which non-linearity, detector
dead-time, and, in the case of $\alpha$\,Cen, possible
contamination by the companion (see Chmielewski et al. 1992 for a
full discussion). Introducing Bessell's color indices into the
Ram\'{\i}rez \& Mel\'endez (2005b) calibrations, along with our
metallicities (Table~\ref{results}), we derive \Teff
($\alpha$\,Cen\,A) = 5794 $\pm$ 34 K and \Teff ($\alpha$\,Cen\,B)
= 5182 $\pm$ 19 K, as a weighted average of the ($B-V$), ($V-R$) and
($V-I$) color indices, the latter two in the Cousins system. These
new photometric \Teffb  determinations, directly from the
calibrations, agree well, for $\alpha$\,Cen\,A, with both the
H$\alpha$ and the spectroscopic one. On the other hand, for
$\alpha$\,Cen\,B this determination lessens slightly, but does not
eliminate, the disagreement between the spectroscopic \Teffb and
the other two. We must therefore state clearly that there is an
offset between the spectroscopic \Teffb of $\alpha$\,Cen\,B and the
other two \Teffb determinations. These last figures, in a classical
spectroscopic analysis of solar-type stars, would be regarded as
``photometric'' \Teffs, to be compared to those obtained from
other methods. Our results are all displayed in
Table~\ref{results}, where we also list the direct surface
gravities resulting from the observed radii and dynamical masses.

Valenti \& Fischer (2005) have derived the atmospheric parameters
of $\alpha$\,Cen\,A and B by means of a different technique. They
have fitted directly large sections of the observed spectra to
synthetic ones, obtaining the atmospheric parameters (this
technique also relies on the excitation \& ionization equilibria
of atomic species). Their analysis is differential with respect to
the Sun, for which they adopted \Teff = 5770, [Fe/H] = $+$0.00 and
log g = 4.44. They quote uncertainties of 44 K, 0.03 dex and 0.06
dex, respectively, for \Teff, log g and [Fe/H]. Their results are
\Teff = 5802 K, log g = 4.33 and [Fe/H] = +0.23, for
$\alpha$\,Cen\,A, and \Teff = 5178 K, log g = 4.56 and [Fe/H] =
+0.21, for $\alpha$\,Cen\,B. These figures are in good agreement,
even within their very small claimed uncertainties, with our
spectroscopic parameters, again, with the exception of the
spectroscopic \Teffb of $\alpha$\,Cen\,B. Particularly, their log g
values are in excellent agreement with the direct log g values of
Table~\ref{results}.

Frutiger et al. (2005) have also analysed the spectra of
$\alpha$\,Cen\,A and B by a fundamentally different and promising
technique. They have inverted high resolution (R $\sim$ 10$^{\rm
5}$), moderately high S/N ($\sim$250) spectra of the stars by
means of a multi-component model photosphere. The components take
into account rotational broadening, center-to-limb variations and
vertical and horizontal flows of surface elements, such as
granules and inter-granular areas. In this approach, the full line
profile is used to constrain the temperature stratification of the
atmosphere, as well as the velocity fields (Allende-Prieto et al.
1998). The technique is rather model-dependent, however, and
should be compared with classical spectroscopic analyses with
caution. For the 3-component models, the ones they favor, \Teffsb
and [Fe/H] substantially lower than those found by us are obtained
(their table 4). For $\alpha$\,Cen\,A, they favor \Teff = 5705 K,
log g = 4.28 $\pm$ 0.03 and [Fe/H] = +0.08 $\pm$ 0.02 (we have
converted their abundances from absolute to relative values with
respect to the Sun, which they also analysed with the same
techniques. Their analysis, in this sense, may also be considered
as differential). For $\alpha$\,Cen\,B, the results are \Teff =
5310 K, log g = 4.74 $\pm$ 0.02 and [Fe/H] = $+$0.05 $\pm$ 0.01.

It is not straightforward to determine the uncertainties of their
\Teffb values, since they only quote the uncertainty of this
parameter as derived by a weighted average of the $\sigma T^4$ 
fluxes of each atmospheric component, weighted by the
filling-factors, by means of the Eddington-Barbier relation
applied to a grey atmosphere. The \Teffsb we quoted are the ones
they computed from the flux spectra obtained from their
temperature stratifications fed into an ATLAS9 model, for which
they provided no formal uncertainty. The uncertainties of their
\Teffsb within the Eddington-Barbier approximation, are,
respectively, 130 K and 67 K for $\alpha$\,Cen\,A and B. Within
these error bars, then, their \Teffsb resulting purely from the
inversion procedure (letting all parameters free) may be regarded
as compatible with ours. Their metallicities, however, are
significantly lower. They also found a very high log g for
$\alpha$\,Cen\,B, prompting them to attempt a larger number of
inversions this for component, first fixing log g = 4.48 dex,
which led to a reduced \Teff = 5154 K with no appreciable change
in metallicity, and then fixing the rotational velocity, which
produced \Teff = 5260 K, again with no significant impact on
metallicity, and a new log g = 4.68 $\pm$ 0.05.

The \Teffsb favored by the inversion method of Frutiger et al.
(2005) add to a complex situation. They seem to be in good
agreement with our spectroscopic \Teffb for $\alpha$\,Cen\,B when
all parameters are independently derived from the inversion
method, but the agreement switches to one with the H$\alpha$ and
photometric \Teffs, when the surface gravity is fixed. All [Fe/H]
values they obtain are lower than ours. For $\alpha$\,Cen\,A,
their surface gravity is in good agreement with ours and the
direct one. It is difficult, however, to reconcile their surface
gravity for $\alpha$\,Cen\,B with our spectroscopic one, and the
direct one given in Table~\ref{results}. It is not clear whether
their results can be directly compared to ours, given the
difference in approach. These authors discuss the possibility of
improving their technique by higher resolution spectra: they
remark on the difficulty of disentangling effects of rotation,
macro-turbulence, granulation and instrumental profile. This
sophisticated approach, probably, can lead to substantially
improved constraints to the atmospheric parameters of solar-type
stars, leading to increased physical insight on the shortcomings
of 1-D, static atmospheric models.

\subsection{Systematic offsets between spectroscopic, Balmer line and photometric \Teff scales}

The study of the atmospheric parameters collected at
Table~\ref{hist-review} reveals an interesting pattern, if one
again considers only the analyses published since the 90s,
invariably based on high S/N spectra acquired with solid-state
detectors. For $\alpha$\,Cen\,A, the works employing the
\ion{Fe}{i}/\ion{Fe}{ii} criterion, alone or combined with another
method, obtained \Teffsb generally higher than those derived
exclusively from photometry (excepting the result of FM90, which
is one of the lowest and applied the excitation \& ionization
approach). Thus, the present work, Santos et al. (2004) and
Neuforge-Verheecke \& Magain (1997) found the highest \Teffs,
while ABLC04, relying only in photometry, derived the lowest
\Teff. Doyle et al. (2005) also used the stellar luminosity and
radius to derive \Teff, and found a value lower than the
spectroscopic ones, but for the FM90 analysis. ABLC04 reported
good agreement of their photometric \Teffsb with those derived from
the H$\beta$ profiles. Interestingly, they also found an offset of
$\sim$120 K between their \Teffsb and the ones of Kovtyukh et al.
(2003), the latter being the higher. Kovtyukh et al. (2003)
derived \Teffsb by the line-depth ration method (Gray \& Johanson
1991), and their \Teffsb were actually calibrated by spectroscopic
ones. The pattern just discussed suggests that the spectroscopic
\Teffb scale is indeed hotter than the photometric and Balmer line
one, and that the latter two are in generally good agreement. The
situation for $\alpha$\,Cen\,B is unfortunately much less clear:
there are less analyses, and the authors employed a more
restricted set of criteria. Nevertheless, the lowest \Teffb is
again due to photometric methods (ABLC04), the highest one in this
case corresponding to the H$\alpha$-derived \Teffb of Chmielewski
et al. (1992).

A disagreement between photometric and spectroscopic \Teffb scales
has been recently pointed out by a number of authors. Ram\'{\i}rez
et al. (2007) discussed how the \ion{Fe}{i}/\ion{Fe}{ii}
ionization equilibrium is not realized in cool stars when an IRFM
\Teffb scale is applied, in the derivation of oxygen abundances
from the $\lambda$7774 triplet lines. Their sample is large, and
they convincingly show (their fig. 5) that the \Teffb offset
between the IRFM and the spectroscopic scales is significant for
\Teffb $\sim$ 5000 K (a reasonable agreement was found for \Teffb
$\sim$ 6000 K). They note that Ram\'{\i}rez \& Mel\'endez (2004)
found the same offset, as did Santos et al. (2004), Yong et al.
(2004) and Heiter \& Luck (2003). Yong et al. (2004) suggested
that non-LTE effects, shortcomings of the model atmosphere
representation for cool stars, or as yet unidentified effects,
might be responsible for the discrepancy. Ram\'{\i}rez et al.
(2007), however, note that Santos et al. (2005) reported good
agreement between IRFM and spectroscopic \Teffb scales. Adding to
this complex picture, Casagrande et al. (2006), in the derivation
of their own IRFM \Teffb scale, also found good agreement between
spectroscopic and IRFM \Teffs. They argue that the disagreement
reported by other authors might be due, at least in part, to
uncertainties in the different absolute flux calibrations adopted,
and suggest that additional direct angular diameter measurements
for a well chosen sample of G and K dwarfs might go a long way
towards clarifying the disagreement of the \Teffb scales.

The results of Casagrande et al. (2006) were essentially backed by
the \Teffb scale of Masana et al. (2006). The latter employed
a variation of the IRFM method, in which the stellar energy
distributions were fitted, from the optical to the IR, to
synthetic photometry computed from stellar atmosphere models. They
found their results only slightly offset from the IRFM results.
The offset, at $\sim$30 K, was deemed to be small, and these
authors optimistically assert that their \Teffb scale agrees with
the spectroscopic one for FGK dwarfs and subgiants, such that
\Teffsb for this class of stars may be regarded as accurate within
$\sim$1\% or better.

The consistency of the different \Teffb scales is sought as an
important confirmation that 1-D, plane-parallel, static and LTE
model atmospheres adequately represent cool stars, if not in the
absolute, at least in the relative sense, provided that the Sun
can be accurately placed in the stellar context. The solar
placement in the \ion{Fe}{i}/\ion{Fe}{ii} excitation \& ionization
equilibria and Balmer line \Teffb scale is obtained by the
observation of solar flux spectra. An accurate photometric
placement of the Sun in the corresponding \Teffb scale, however, is
a difficult task still beset with large uncertainties (see, e.g.,
Holmberg et al. 2006 for an up-to-date discussion). These three
\Teffb scales actually gauge rather dissimilar physical quantities.
The excitation \& ionization \Teffb is obtained by matching models
to observed spectral line intensities. The Balmer line \Teffb
measures the temperature stratification of the atmosphere, which
is mapped onto the line wings by the depth-dependence of the
source-function. The photometric \Teffb must reproduce the stellar
flux distribution in a large wavelength regime, and is the one
most directly tied to the fundamental definition of effective
temperature (B\"ohm-Vitense 1981). As long as the consistency
between these scales is realized no better than within $\sim$150
K, the \Teffb of cool dwarfs and subgiants will remain uncertain by
this amount at the very least.

Non-LTE and other possibly more complex effects have repeatedly
been blamed for offsets between spectroscopically and
photometrically derived atmospheric parameters in cool stars.
Schuler et al (2006a), in their analysis of Hyades dwarfs,
reported a systematic offset of the oxygen abundances derived from
the $\lambda$7774 triplet lines, for \Teffb $<$ 5450 K, in the
opposite sense of the NLTE expectations. They tentatively suggest
that chromospheric activity might be at least partially
responsible for the offset, an explanation also concurred by
Morell \& Micela (2004), though the latter propose that model
atmosphere pitfalls might be also present. Schuler et al. (2006b)
reinforce this interpretation in an analysis of the $\lambda$6300
[OI] line in the very active Hyades stars, reporting offsets
between \ion{Fe}{i}/\ion{Fe}{ii} abundances which increase as
\Teffb $\leq$ 5000 K. It should be emphasized that high
chromospheric activity is unlikely to be a source of the \Teffb
discrepancy of $\alpha$\,Cen\,B, since both components of the
system are inactive stars, which probably implies that the
problem is more complex. We also draw attention to the result of 
Shchukina \& Trujillo-Bueno (2001), who found an offset of
the \ion{Fe}{i}/\ion{Fe}{ii} abundances of the Sun. They interpret
this offset as well explained by NLTE effects amounting to 0.07
dex for \ion{Fe}{i}, the best fit Fe abundance for a LTE analysis
being the lower by this amount. They assert that a full 3-D, NLTE
model atmosphere formulation is able to bring the solar
photospheric Fe abundance in line with the meteoritic one, at log
N(Fe) = 7.50 (in the usual scale where log N(H) = 12.00). The main
cause of the offset is the overionization of \ion{Fe}{i}, and the
larger errors are seen in the \Wlamsb of low excitation lines,
which are weaker in the NLTE case.

It is interesting to note that this effect, in a classical LTE
model atmosphere analysis, would result in the overabundance of
the high excitation \ion{Fe}{i} lines, an effect naturally
interpreted as too low a \Teffb being attributed to the model. This
is exactly the condition necessary for a 1-D, LTE analysis to lead
to the high spectroscopic \Teffb $-$ that we obtained. Forcing the
\ion{Fe}{i}/\ion{Fe}{ii} abundances into agreement in a LTE
analysis would indeed call for a higher \Teffb, to a first
approximation, by $\sim$100 K, a value similar to the difference
between our spectroscopic and H$\alpha$/photometric \Teffsb of
$\alpha$\,Cen\,B. We suggest that the similarity in atmospheric
parameters between $\alpha$\,Cen\,A and the Sun, in the context of
a differential analysis, led to a good agreement between the three
\Teffb criteria for the former. For $\alpha$\,Cen\,B, a much cooler
object, an imperfect cancellation of the presence of NLTE effects
is probably the reason why the three different \Teffb criteria do
not agree.

Even if the presence of NLTE effects and other problems can be
established, one must keep in mind that other uncertainties are
present in the photometric and H$\alpha$ \Teffb scales, as
discussed above. It would be very valuable to extend the novel
approaches of Valenti \& Fischer (2005) and Frutiger et al. (2005)
to additional stars for which stringent observational constraints
are available, to quantify how discrepancies of the type discussed
here do occur for objects with atmospheric parameters increasingly
different from the Sun's. At present, we can state that it is
likely that our spectroscopic \Teffb for $\alpha$\,Cen\,B is more
uncertain that the other two determinations, being too high. This
conclusion is, however, drawn in the context of a confusing
picture. Additional work is clearly necessary before a definitive
judgment can be passed on the consistency between the excitation
\& ionization,  Balmer line and photometric \Teffb scales can be
reached.

Notwithstanding the discrepancies, if we follow an usual
practice in spectroscopic abundance analysis which obtain
atmospheric parameters with more than one criterion, mean \Teffsb
for the $\alpha$\,Cen\ stars can be calculated from the values of
Table~\ref{results}, averaged by their inverse variances. The
results are \Teffb ($\alpha$\,Cen\,A) = 5824 $\pm$ 27 K and \Teffb
($\alpha$\,Cen\,B) = 5223 $\pm$ 62 K, where the quoted
uncertainties are the standard deviations of the average, and do
not reflect external and systematic errors. Good agreement between
our ionization surface gravities, and those directly determined
from measured masses and radii, within the errors, for both stars,
is realized.

\begin{table*}
\caption[]{The atmospheric parameters resulting from our
spectroscopic analysis: from the excitation \& ionization
equilibria of \ion{Fe}{i} and \ion{Fe}{ii} (excitation \Teff,
ionization $\log g$, $[$Fe/H$]$ and $\xi_t$), from the wings of
H$\alpha$ and the photometric calibrations of Ram\'{\i}rez \&
Mel\'endez (2005b)(photometric). The mean \Teffb is  weighted by
the inverse variances. The ``direct'' $\log g$ values are derived
from the masses of Pourbaix et al. (2002) and the radii of
Kervella et al. (2003) (see text).} \label{results}
\begin{center}
\begin{tabular}{l c c c c c c c c c}\hline
&\Teff(K)&\Teff(K)&\Teff(K)&\Teff(K)&$\log g$&$\log g$&\multirow{2}{*}{$\rm
[Fe/H]$}&\multirow{2}{*}{$\xi_t$(km\,s$^{-1}$)}\\
                &excitation       & H$\alpha$      & photometric   &
weighted mean  &ionization& direct&      &\\\hline
{$\alpha$\,Cen\,A}&\tpm{5847}{27}&\tpm{5793}{50}   &
\tpm{5794}{34}
&\tpm{5824}{26}&\tpm{4.34}{0.12}&\tpm{4.307}{0.005}&$+$\tpm{0.24}{0.03}&\tpm{1.46}{0.03}\\
{$\alpha$\,Cen\,B}&\tpm{5316}{28}&\tpm{5155}{100}  &
\tpm{5182}{19}
&\tpm{5223}{62}&\tpm{4.44}{0.15}&\tpm{4.538}{0.008}&$+$\tpm{0.25}{0.04}&\tpm{1.28}{0.12}\\\hline
\end{tabular}
\end{center}
\end{table*}

\begin{figure*}
\begin{center}
\resizebox{8.5cm}{!}{\includegraphics{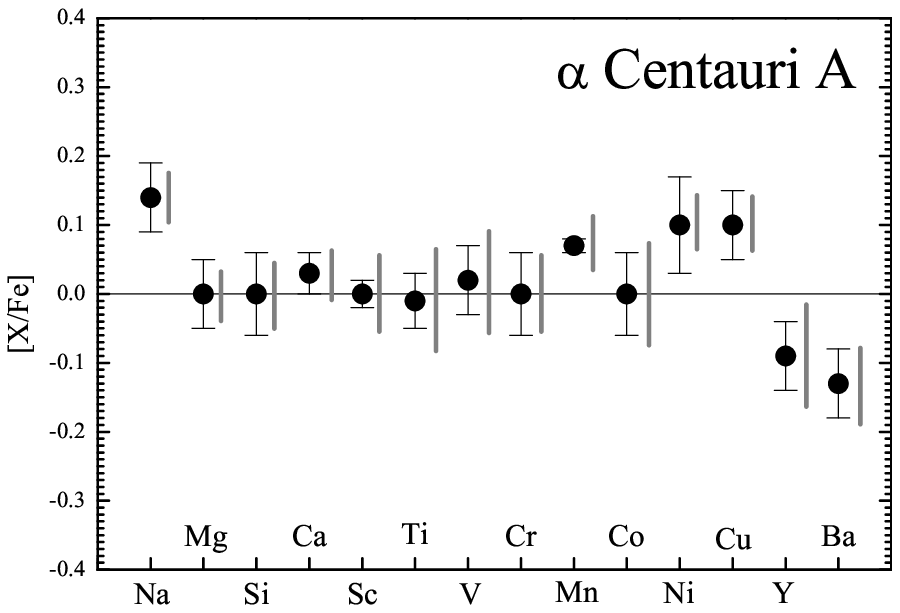}}
\resizebox{8.5cm}{!}{\includegraphics{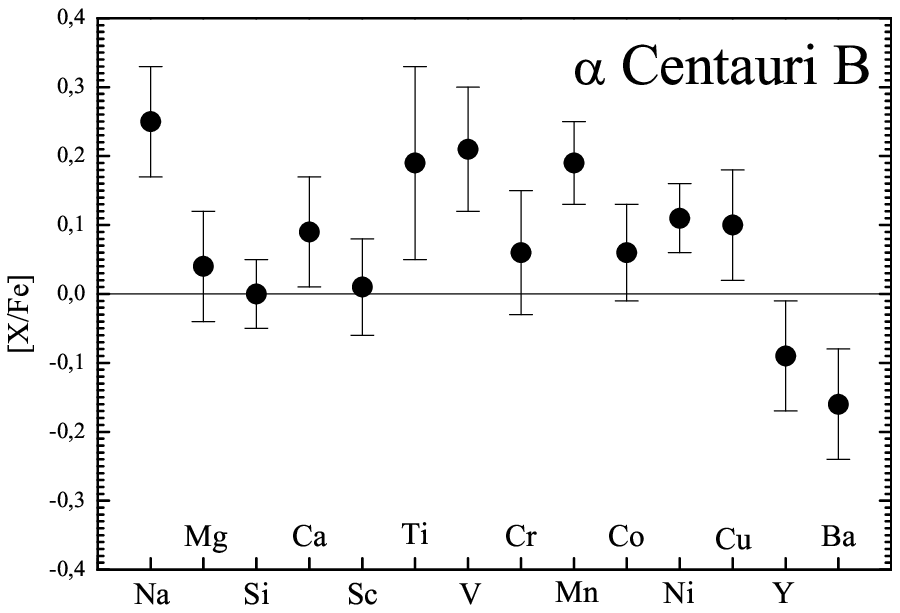}}
\end{center}
\caption[]{The abundance pattern of $\alpha$\,Cen\,A \& B. 
The uncertainty bars are the dispersions of the abundances given
by the element lines. The gray bars beside the data points in the
$\alpha$\,Cen\,A plot correspond to the total compounded errors
arising from the atmospheric parameters and the \Wlamsb (see
Table~\ref{abunds}). Na is seen to be overabundant, while a solar
pattern is seen from Mg to Co, but for an excess of Mn. Ni and Cu
are also overabundant. Some doubt can be cast about Ti and V,
since they seem overabundant in $\alpha$\,Cen\,B although solar in
$\alpha$\,Cen\,A. The slow neutron capture elements Y and Ba are
in clear deficit. The bigger uncertainty bars seen in
$\alpha$\,Cen\,B are probably a result of a less accurate
normalization of the spectra of a cooler star, and may also
be due to its less accurate atmospheric parameters.}
\label{abundance}
\end{figure*}

\begin{figure*}
\begin{center}
\resizebox{15cm}{!}{\includegraphics{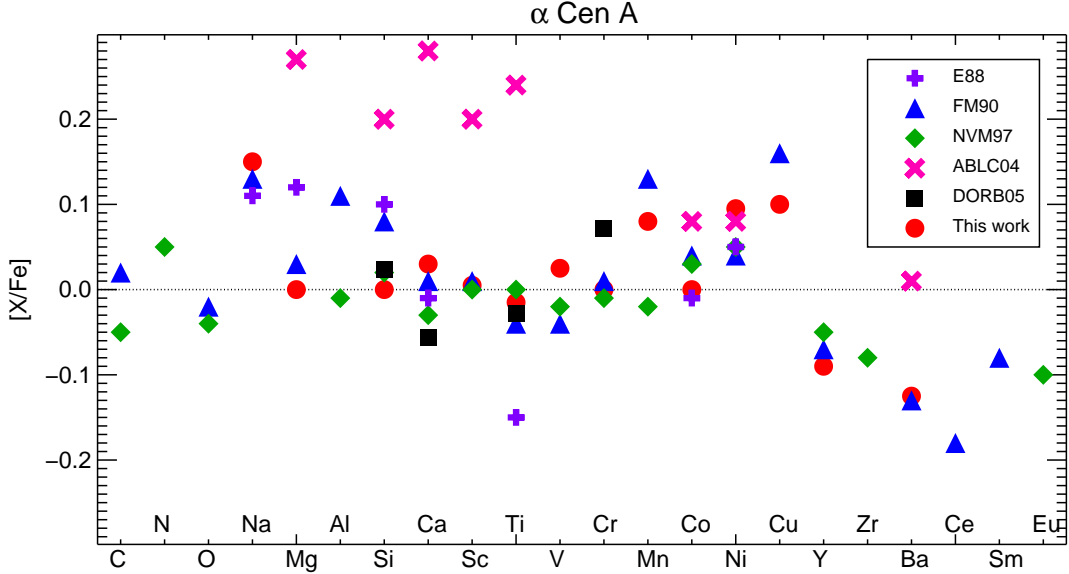}}
\end{center}
\caption[]{A comparison of the $[$X/Fe$]$ abundance pattern of
$\alpha$\,Cen\,A derived by different authors. E88 stands for
Edvardsson (1998). Only elements for which abundances are based on
more than one line are included. Apart from the ABLC04 results,
which point to excesses of nearly all elements with respect to the
Sun, most published data imply an abundance pattern in which C, N,
O, Ca, Sc, V and Cr are normal; Na, Mn, Co, Ni and Cu are
enhanced; Mg, Al and Si are possibly enhanced; and Ti and all
elements heavier than Y are under-abundant.}
\label{abucompar1}
\end{figure*}

\begin{figure*}
\begin{center}
\resizebox{15cm}{!}{\includegraphics{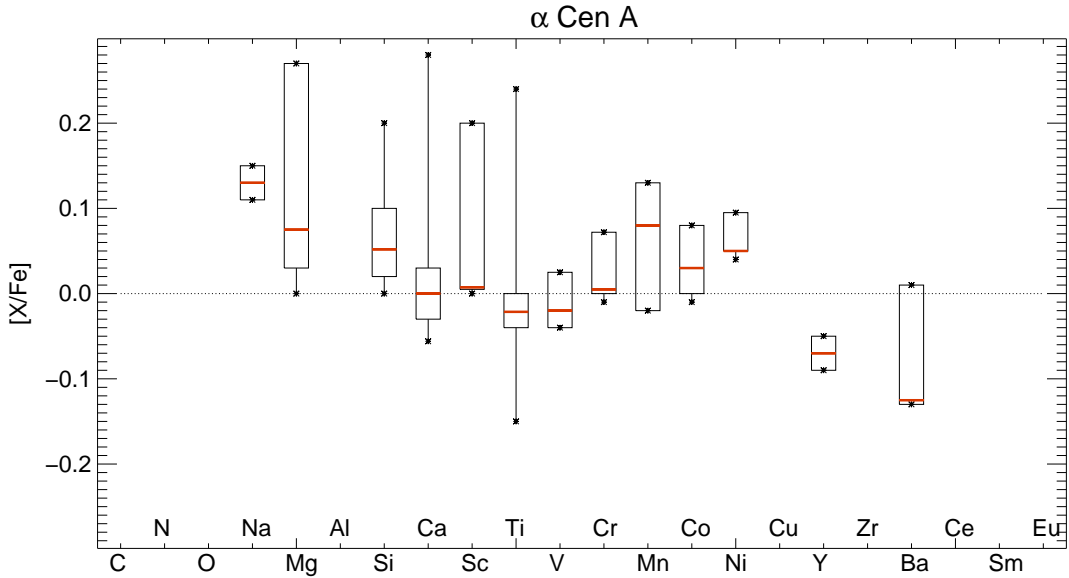}}
\end{center}
\caption[]{The same as Fig.\ref{abucompar2}, but averaging the
$[$X/Fe$]$ abundances of the elements with published data from at
least three authors. The boxes comprise 50\% of the data points,
and are centered at the mean. The horizontal dashes inside the
boxes mark the median. The whiskers mark the maximum and minimum
values. In this more stringent statistical analysis of the
available data, Na, Mg, Si, Mn, Co and Ni are enhanced; Ca, Sc,
Ti, V and Cr are normal; Y and Ba are under-abundant.}
\label{abucompar2}
\end{figure*}

\begin{table}
\caption[]{The abundance pattern of the $\alpha$\,Cen stars. The
first two columns give the [X/Fe] ratios with the corresponding
observed dispersions of the line abundances, for those elements
with three or more available lines, for $\alpha$\,Cen\,A. The next
two columns give the corresponding data for $\alpha$\,Cen\,B. the
fifth column provides the total uncertainty of the [X/Fe] ratios
corresponding to errors in \Teff, log g, [Fe/H], $\xi$ and \Wlam,
respectively, of 30 K, 0.12 dex, 0.04 dex, 0.03 km\,s$^{-1}$ and
2.9 m\AA, composed in quadrature (the latter enters twice, see
section 3).} \label{abunds}
\begin{center}
\begin{tabular}{l r c c r c c r c}\hline
\multirow{2}{*}{element}&&\multicolumn{2}{c}{\acena}&&\multicolumn{2}{c}{\acenb}&&\multirow{2}{*}{$\Delta$}\\
                                                                                                       \cline{3-4}
                                                                                                       \cline{6-7}
          && [X/Fe]  &$\sigma$        &&[X/Fe]   & $\sigma$       &&             \\\hline
Na        &&$+$0.14  & ...            &&$+$0.25  & ...        &&0.03             \\
Mg        &&\mezero  & ...            &&$+$0.04  & ...        &&0.03             \\
Si        &&\mezero  &0.06            &&\mezero  &0.05        &&0.05             \\
Ca        &&$+$0.03  &0.03            &&$+$0.09  &0.08        &&0.03             \\
Sc        &&\mezero  &0.02            &&$+$0.01  &0.07        &&0.06             \\
Ti        &&$-$0.01  &0.04            &&$+$0.19  &0.14        &&0.06             \\
V         &&$+$0.02  &0.05            &&$+$0.21  &0.09        &&0.07             \\
Cr        &&\mezero  &0.06            &&$+$0.06  &0.09        &&0.06             \\
Mn        &&$+$0.07  &0.01            &&$+$0.19  &0.06        &&0.04             \\
Co        &&\mezero  &0.06            &&$+$0.06  &0.07        &&0.07             \\
Ni        &&$+$0.10  &0.07            &&$+$0.11  &0.05        &&0.04             \\
Cu        &&$+$0.10  & ...            &&$+$0.10  & ...        &&0.04             \\
Y         &&$-$0.09  &0.05            &&$-$0.09  &0.08        &&0.07             \\
Ba        &&$-$0.13  &0.05            &&$-$0.16  &0.08 &&0.06
\\\hline
\end{tabular}
\end{center}
\end{table}

\section{Abundance pattern}

The abundances of the other elements were obtained with the
adopted atmospheric model of each star, corresponding to the
spectroscopic \Teffs, and the log g, $[$Fe/H$]$ and $\xi$ values
as given in Table~\ref{results}. Average abundances were
calculated by the straight mean of the individual line abundances.
For Sc, Ti and Cr, good agreement, within the errors, was obtained
for the abundances of the neutral and singly ionized species, and
thus these species confirm the \ion{Fe}{i}/\ion{Fe}{ii} ionization
equilibria. The results are plotted in Fig.~\ref{abundance} as
[X/Fe] relative to the Sun. In this figure the error bars are
merely the internal dispersion of the individual line abundances.
For \ion{Na}{i}, \ion{Mg}{i}, \ion{Cu}{i} and \ion{Ba}{ii}, the
dispersions refer to the difference between the abundances of the
two available lines for each element. There is a very good
consistency between the abundance patterns of the two stars, but
for the the \ion{Ti}{i} and \ion{V}{i} abundances. The larger
error bars for $\alpha$\,Cen\,B, however, lead us to consider the
two abundance patterns as consistent with each other.

In Fig.~\ref{abundance}, the vertical dark grey bars besides the
data points of the abundance pattern of $\alpha$\,Cen\,A are the
composed {\it r.m.s.} uncertainties, for each element, calculated
by varying the spectroscopic atmospheric parameters of
$\alpha$\,Cen\,A by the corresponding uncertainties of
Table~\ref{results}. To this calculation we added the abundance
variations caused by summing to all \Wlamsb the 2.9 m{\AA}
uncertainty of the correction of Fig~\ref{gauss-voigt}. This
\Wlamb uncertainty enters twice: once for the uncertainty in the
corrected moon \Wlams, reflecting onto the solar log $gf$s, and
another one due to stellar \Wlamb themselves. In
Fig~\ref{abundance}, it is apparent that the abundance variations
due to the uncertainties in the atmospheric parameters and \Wlamsb
are comparable to the observed dispersions of the line abundances
for $\alpha$\,Cen\,A. For $\alpha$\,Cen\,B, the line abundance
dispersions are generally larger, probably due to its more
uncertain \Wlams, but also, as discussed above, possibly
owing to its larger \Teffb difference from the standard object (the
Sun) and its more uncertain atmospheric parameters.

Our abundance pattern for $\alpha$\,Cen\,A is clearly the most
reliable of the pair, and is directly compared to those of other
authors in Fig~\ref{abucompar1}. Only abundances represented by
more than one spectral line are shown. The observed dispersion is
comparable to the uncertainties normally quoted in spectroscopic
analyses. Only for the light elements between Mg and Ti is a
larger disagreement observed, in this case due to the analysis of
ABLC04, in which abundances are higher than in the bulk of other
data by $\sim$0.2 dex. For the elements heavier than V,
essentially all data agree that V and Cr have normal abundance
ratios, that Mn, Co, Ni and Cu are enhanced, and all heavy
elements from Y to Eu are deficient in the abundance pattern of
$\alpha$\,Cen\,A with respect to the Sun, with the sole exception
of Ba, for which ABLC04 found a normal abundance. The available 
literature data also suggests that the C, N and O abundance
ratios of $\alpha$\,Cen\,A are solar.

This statistical analysis can be extended if we regard only the
elements for which at least three independent studies provided
data. This is shown in Fig~\ref{abucompar2}, for a more select
sample of elements. We may conclude, with somewhat greater
robustness, considering the number of abundance results, that Na,
Mg, Si, Mn, Co and Ni are over-abundant; that Ca, Sc, Ti, V and Cr
have solar abundance ratios; and that Y and Ba are over-deficient
in the abundance pattern of $\alpha$\,Cen\, with respect to the
Sun.

The high metallicity of the $\alpha$\,Cen\ system, and its space
velocity components (U,V,W)(km.s$^{\rm -1}$) = (-24, +10, +8)
(Porto de Mello et al. 2006, all with respect to the Sun) place it
unambiguously as a thin disk star. We next analyze its abundance
ratios, for the elements with more reliable data, as compared to
recent literature results for metal-rich stars. Bensby et al.
(2003, hereafter BFL) analysed 66 stars belonging to the thin and
thick disks of the Milky Way, deriving abundances of Na, Mg, Al,
Si, Ca, Ti, Cr, Fe, Ni and Zn. Bodaghee et al. (2003, hereafter
BSIM) analysed a sample of 119 stars, of which 77 are known to
harbor planetary companions, deriving abundances of Si, Ca, Sc,
Ti, V, Cr, Mn, Fe, Co and Ni. The data of the latter study comes
from the Geneva observatory planet-search campaign (e.g., Santos
et al. 2005). Since they concluded that planet-bearing stars
merely represent the high metallicity extension of the abundance
distribution of nearby stars, their full sample can be used to
adequately represent the abundance ratios of metal rich stars,
without distinction to the presence or absence of low mass
companions. These two works have in common the important fact that
they sample well the metallicity interval $+$0.20 $<$ $[$Fe/H$]$
$<$ $+$0.40, an essential feature for our aim.

The elements in common between the two abundance sets are
Si, Ca, Ti, Cr and Ni. It can be concluded (Fig. 13 of BFL, Fig. 2
of BSIM) that, in $[$Fe/H$]$ $\geq$ $+$0.20 stars, Ca is
under-abundant, Ti is normal and Ni is enhanced. For Si, BFL
suggest abundance ratios higher than solar, while BSIM found a
normal abundance. For Cr, the data of BSIM suggests abundance
ratios lower than solar, while BFL found solar ratios. For the
elements not in common in the two studies, Na, Mg, Al, Sc, V, Mn
and Co are found to be enhanced in $[$Fe/H$]$ $\geq$ $+$0.20
stars, while Zn has normal abundance ratios.

An interesting feature of the $\alpha$\,Cen\ abundance pattern is
the under-abundance in the elements heavier than Y, which could be
reliably established for Y and Ba (Fig~\ref{abucompar2}). This
result is confirmed by Bensby et al. (2005), who found both for Y
and Ba lower than solar abundance ratios for metal rich thin disk
stars. Another interesting feature, the excess of Cu (found by us
and FM90), can be checked with the recent results of Ecuvillon et
al. (2004), who also obtain for $[$Fe/H$]$ $\geq$ $+$0.20 stars an
average $[$Cu/Fe$]$ $\sim$0.1 dex.

Merging our evaluation of the abundance pattern of
$\alpha$\,Cen\,A, from the available independent analyses, with
the previous discussion, we conclude that $\alpha$\,Cen\,A is a
normal metal-rich star in its Na, Mg, Si, Ti, V, Cr, Mn, Co and Ni
abundances. The result for Ca is inconclusive, and only for Sc
does its abundance diverge from the BFL/BSIM data, in that its
normal abundance ratio contrasts with the overabundance found for
$[$Fe/H$]$ $\geq$ $+$0.20 stars by BSIM. It seems reasonably well
established then that the $\alpha$\,Cen\ system is composed of two
normal metal rich stars when regarded in the local disk
population.

\section{Evolutionary state}

\begin{figure}
\begin{center}
\resizebox{9cm}{!}{\includegraphics{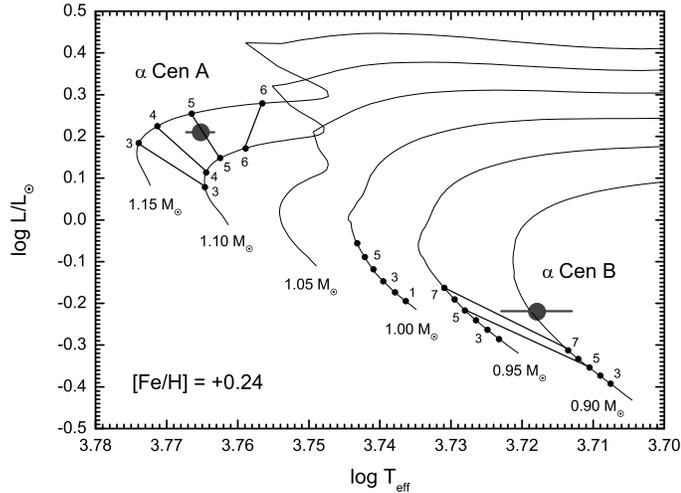}}
\end{center}
\caption[]{The evolutionary state of the $\alpha$\,Cen system. The
stars are plotted in the HR diagram superimposed to the isochrones
and evolutionary tracks of Kim et al. (2002) and Yi et al. (2003).
The horizontal error bars refer to the uncertainties of
Table~\ref{results}. The actual uncertainties in the luminosities
are smaller than the symbol size. The masses derived from
evolutionary tracks, M$_{\rm A}$ = 1.13 $\pm$ 0.01 and M$_{\rm B}$
= 0.89 $\pm$ 0.05, agree very well with the orbital solution of
Pourbaix et al. (2002). The tracks are labelled with masses in
solar units. the numbers alongside the tracks are ages in Gyr. The
thin solid lines between the tracks join points with the same age.
For $\alpha$Cen\,A, an age of 4.5 to 5.3 Gyr can be inferred. A
match of $\alpha$Cen\,B within this age range is only possible if
its \Teffb is revised upwards (see text).} \label{ages}
\end{figure}

An important outcome of the present analysis is to establish
if the derived atmospheric parameters, coupled to high-quality
parallaxes, allow a consistent determination of the masses and
ages of the $\alpha$\,Cen system in a traditional HR diagram
analysis, matching the stringent constraints posed by the
orbital solution and seismological data. In Fig.~\ref{ages} we
plot the position of the $\alpha$\,Cen components in the
theoretical HR diagram of Kim et al. (2002) and Yi et al. (2003),
corresponding to its exact metallicity and a solar abundance
pattern. The bolometric corrections were taken from Flower (1996),
and in calculating the luminosities the Hipparcos (ESA 1997)
parallaxes and visual magnitudes were used. Evolutionary tracks
and isochrones from different authors (e.g., Girardi et al 2000,
Charbonnel et al. 1999, Schaller et al. 1992) were also tested,
and good agreement between the different tracks was found, to
better than $\sim$50 K, for the position of both $\alpha$\,Cen
components. This is an expected result, given that the Sun is
generally used to calibrate these models. The solar mass, radius
and age provides a zero point to the models and allow a solution
as a function of the adopted mixing-length of the convection
theory (still a free parameter) and the initial helium abundance.
Thus, differences between the models can be substantial in the
treatment of stars which are very different from the Sun (e.g., 
Lyra et al. 2006), but good agreement for solar-type stars is 
a natural outcome of this procedure. The conclusions drawn below, 
then, are essentially model-independent, at least for the \Teffb 
and luminosity intervals involved here.

The \Teffb values and error bars in the diagram are those of the
weighted mean of table~\ref{results}. From the diagram, masses of
M$_{\rm A}$ = 1.13 $\pm$ 0.01 and M$_{\rm B}$ = 0.89 $\pm$ 0.03
can be derived, and agree well with the orbital solution of
Pourbaix et al. (2002). The age of $\alpha$\,Cen\,A can be
relatively well constrained to the interval of 4.5 to 5.3 Gyr
(1$\sigma$), again, in good agreement with the seismological
results of Y{\i}ld{\i}z (2007), Eggenberger et al. (2004), Miglio \&
Montalb\'an (2005) and Thoul et al. (2003), within the quoted
uncertainties. We conclude that, adopting the average \Teffsb and
[Fe/H] found in this work, the position of $\alpha$\,Cen\,A in
up-to-date theoretical HR diagrams can be reconciled both with
seismological and dynamical data. Despite its higher mass and its
being (probably) older than the Sun, the higher metallicity slows
the evolution to the point that the star has not yet reached the
``hook'' zone of the HR diagram, thus enabling a unique age
solution through this type of analysis.

Concerning $\alpha$\,Cen\,B, it is also apparent in Fig~\ref{ages}
that its position cannot be reconciled, within 1$\sigma$, with an
age near 5 Gyr, as was possible for $\alpha$\,Cen\,A. However, an
upward revision of only $\sim$60 K would bring its position in
agreement with a track of 0.93 solar mass (the dynamical value),
and an age of $\sim$5-8 Gyr. Given the uncertainties discussed in
section 3, along with the probable effect of systematic errors in
the \Teffb determination of $\alpha$\,Cen\,B, this is not outside
the 2$\sigma$ confidence interval of the results. We conclude that
$\alpha$\,Cen\,A has its position in the theoretical HR diagram
well matched by up-to-date models, within the uncertainties of the
determination of its atmospheric parameters, and also within the
small differences, in this \Teffb and luminosity regime, between
different grids of evolutionary models. $\alpha$\,Cen\,B, in its
turn, has been shown to suffer from a larger uncertainty in its
\Teff, precluding a more stringent assessment of a match between
its evolutionary mass and age and the results from the dynamical
solution and asteroseismology. Further data are still necessary to
allow a more definitive conclusion on this issue. If indeed its
spectroscopic \Teffb is systematically offset, the mean \Teffb we
derived would decrease and displace its position on the HR diagram
to the right, forestalling a match with the age of
$\alpha$\,Cen\,A. For a better understanding of  the onset of
possible NLTE effects in cool stars, and the hindrance thus
incurred in the determination of their atmospheric parameters, it
would be interesting to perform further analyses of such objects,
for which high-quality spectra could be secured, and the relevant
observational constraints made at least partially available.
Interesting bright, nearby K-dwarf candidates for such an
enterprise are $\epsilon$\,Eri, 36 and 70\,Oph, o$^{\rm 2}$ Eri
and $\sigma$ Dra.

\section{Conclusions}

We have undertaken a new detailed spectroscopic analysis of the
two components of the $\alpha$ Centauri binary system, and have
attempted an appraisal of the many discordant determinations of
its atmospheric parameters and abundance pattern, and of the
sources of errors in their determination. We derived purely
spectroscopic atmospheric parameters, from R = 35\,000 and S/N $>$
1\,000 spectra, in a strictly differential analysis with the Sun
as the standard. We obtained \Teff$_{\rm A}$ = 5847 $\pm$ 27 K and
\Teff$_{\rm B}$ = 5316 $\pm$ 28 K from the spectroscopic analysis,
and \Teff$_{\rm A}$ = 5824 $\pm$ 26 K and \Teff$_{\rm B}$ = 5223
$\pm$ 62 K from the average of spectroscopic, H$\alpha$ and
photometric \Teffs). We derived $[$Fe/H$]$ = $+$0.24 $\pm$ 0.03
dex for the system. The spectroscopic surface gravities, log
g~$_{\rm A}$ = 4.34 $\pm$ 0.12 and log g~$_{\rm B}$ = 4.44 $\pm$
0.15, are a good match to those determined from directly measured
masses and radii. Good agreement, in both components, is found
between the photometric \Teffb and the one resulting from the
fitting of the wings of H$\alpha$. For $\alpha$\,Cen\,A, these two
\Teffsb also agree with the spectroscopic one. However, for
$\alpha$\,Cen\,B, the spectroscopic \Teffb was found to be
significantly higher, by $\sim$140 K, than the other two. A
comparison of the published \Teffsb for the system in the last 20
years roughly support a spectroscopic \Teffb scale hotter than the
ones owed to photometric methods or the fitting of Balmer lines.

A comparison with recent results from other techniques revealed an
unclear picture. Atmospheric parameters for the $\alpha$\,Cen
stars derived by Valenti \& Fisher (2005) by fitting
directly synthetic spectra to a large spectral coverage, agree
well with our determinations, but for the spectroscopic \Teffb of
$\alpha$\,Cen\,B. Their surface gravities and metallicities are
also in line with our figures. Frutiger et al. (2005), inverting
high-resolution line profiles, found a substantially lower [Fe/H],
and their model-dependent \Teffsb agree either with our
spectroscopic or with the photometric/H$\alpha$ \Teffb, depending
on assumptions. Their log g for $\alpha$\,Cen\,B is also higher
than all other recent determinations.

We discuss possible origins of the offset between the \Teffb
scales, concluding that the presence of NLTE effects, and
also a possible inconsistency between spectroscopic and
photometric \Teffb scales, are probable explanations. Recent
results reporting offsets between spectroscopic and photometric
\Teffb scales in cool stars, of similar magnitude, lend some
credence to this interpretation. But we note that some authors
claim consistency between the two scales, and that other sources
of errors may be at play, such as uncertainties in the absolute
flux calibration of photometric \Teffs. We also note that
recent claims of such \Teffb offsets as caused by chromospheric
activity cannot explain the present discrepancy given that both
$\alpha$\,Cen stars are considerably inactive stars. These
discordant data still preclude a clear evaluation of the problem.
For both $\alpha$\,Cen\,A and B, the spectroscopic surface
gravities agree well, within the uncertainties, with direct values
derived from the dynamical masses of Pourbaix et al. (2002) and
the radii of Kervella et al. (2003). The atmospheric parameters
resulting from our analysis are collected in Tab~\ref{results}.

The abundance pattern of the system, when the various authors's
data are considered for those elements for which at least three
independent analyses are available, is found to be enriched in Na,
Mg, Si, Mn, Co, Ni and (with less reliability) Cu, and deficient
in Y and Ba (Fig~\ref{abucompar2}). This abundance pattern is
found to be in very good agreement with recent results on the
abundance ratios of metal-rich stars. Thus, the system may be
considered as a normal pair of middle-aged, metal-rich, thin disk
stars.

An analysis of the evolutionary state of the system in the
theoretical tracks of Kim et al. (2002) and Yi et al. (2003)
yields a very good agreement of the evolutionary mass (M$_{\rm A}$
= 1.13 $\pm$ 0.01) and age (4.5-5.3 Gyr) of $\alpha$\,Cen\,A with
the results of recent seismological and dynamical data
(Fig~\ref{ages}). For $\alpha$\,Cen\,B, a 1$\sigma$ upward
revision of its \Teffb would bring its position in the HR diagram
within reasonable agreement with the age found for
$\alpha$\,Cen\,A, and an evolutionary mass (M$_{\rm B}$ $\sim$
0.93) in good agreement with the dynamical one would result. This
merely marginal compatibility suggests that, in order to fulfill
the privileged situation of the $\alpha$\,Cen system as a
fundamental calibrator of the modelling of stellar structure and
atmosphere models, additional analyses of component B seem to be
necessary to quantify the onset and magnitude of possible 
NLTE in cool stars, as well as allow a more precise evaluation of
possible offsets between spectroscopic and photometric \Teffb
scales in this class of objects.

\begin{acknowledgements}

We acknowledge fruitful discussions with Bengt Edvardsson. W. L.
and G. R. K. wish to thank CNPq-Brazil for the award of a
scholarship. G. F. P. M. acknowledges financial support by CNPq
grant n$^{\circ}$ 476909/2006-6, FAPERJ grant n$^{\circ}$
APQ1/26/170.687/2004, and a CAPES post-doctoral fellowship
n$^{\circ}$ BEX 4261/07-0. We thank the staff of the OPD/LNA for
considerable support in the observing runs necessary for this
project. Use was made of the Simbad database, operated at CDS,
Strasbourg, France, and of NASA's Astrophysics Data System
Bibliographic Services. Criticism and suggestions from the
anonymous referee have considerably improved the manuscript.

\end{acknowledgements}

\end{document}